\documentclass[12pt,a4paper]{article}
\usepackage{amsmath,amssymb,hyperref,bm,bbm}
\usepackage[utf8]{inputenc}
\numberwithin{equation}{section}
\allowdisplaybreaks[3]

\begin{document}
\begin{titlepage}

\renewcommand{\thefootnote}{\fnsymbol{footnote}}
\begin{flushright}
\begin{tabular}{l}
YITP-20-137\\
\end{tabular}
\end{flushright}

\vfill
\begin{center}


\noindent{\large \textbf{Higher rank FZZ-dualities}}



\vspace{1.5cm}

\noindent{Thomas Creutzig$^{a}$\footnote{E-mail: creutzig@ualberta.ca}
and Yasuaki Hikida$^b$\footnote{E-mail: yhikida@yukawa.kyoto-u.ac.jp}}
\bigskip

\vskip .6 truecm
\centerline{\it $^a$Department of Mathematical and Statistical Sciences, University of Alberta,} \centerline{\it Edmonton, Alberta T6G 2G1, Canada}
\medskip
\centerline{\it $^b$Center for Gravitational Physics, Yukawa Institute for Theoretical Physics,}
\centerline{\it  Kyoto University, Kyoto 606-8502, Japan}

\end{center}
		
\vfill
\vskip 0.5 truecm
\begin{abstract}

We propose new strong/weak dualities in two dimensional conformal field theories by generalizing the Fateev-Zamolodchikov-Zamolodchikov (FZZ-)duality between Witten's cigar model described by the $\mathfrak{sl}(2)/\mathfrak{u}(1)$ coset and sine-Liouville theory. In a previous work, a proof of the FZZ-duality was provided by applying the reduction method from $\mathfrak{sl}(2)$ Wess-Zumino-Novikov-Witten model to Liouville field theory and the self-duality of Liouville field theory. In this paper, we work with the coset model of the type $\mathfrak{sl}(N+1)/(\mathfrak{sl}(N) \times \mathfrak{u}(1))$ and propose that the model is dual to a theory with an $\mathfrak{sl}(N+1|N)$ structure. We derive the duality explicitly for $N=2,3$ by applying recent works on the reduction method extended for $\mathfrak{sl}(N)$ and the self-duality of Toda field theory.
Our results can be regarded as a conformal field theoretic derivation of the duality of the Gaiotto-Rap\v{c}\'ak corner vertex operator algebras 
$Y_{0, N, N+1}[\psi]$ and $Y_{N, 0, N+1}[\psi^{-1}]$.

\end{abstract}
\vfill
\vskip 0.5 truecm

\setcounter{footnote}{0}
\renewcommand{\thefootnote}{\arabic{footnote}}
\end{titlepage}

\newpage

\tableofcontents

\section{Introduction}

In this paper, we propose new strong/weak dualities in two dimensional conformal field theories (CFTs) and explicitly derive simple cases. In general, strong/weak dualities are quite useful, for instance, to examine strongly coupled physics from weakly coupled theories. A rare solvable example was proposed by Fateev-Zamolodchikov-Zamolodchikov (FZZ) \cite{FZZ}, and their duality states an equivalence between Witten's cigar model \cite{Witten:1991yr} and so-called sine-Liouville theory. The cigar model can be described by the coset
\begin{align}
\frac{\mathfrak{sl}(2)_k}{\mathfrak{u}(1)} \, . \label{sl2coset}
\end{align}
The level $k$ is related to the inverse of curvature and the sigma model description is suitable for large $k$.
On the other hand, the sine-Liouville theory is described by a bosonic field $\phi$ with background charge and another bosonic field $X$ with $X \sim X + 2 \pi \sqrt{k}$. The theory includes interaction terms
\begin{align}
V_\text{int} = 2 \pi \lambda  ( e^{ \phi/ b_{(2)} + i \sqrt{k}  \tilde X }  + e^{ \phi/ b_{(2)} - i \sqrt{k}  \tilde X } )\, , \label{sineLiouville}
\end{align}
where $b_{(2)} = 1/\sqrt{k-2}$ and $\tilde X$ represents the dual of $X$. The interaction terms do not grow rapidly when $b_{(2)}$ is large enough.  From this, we can see that the FZZ-duality is an example of strong/weak duality.
See \cite{Kazakov:2000pm} for more details of the FZZ-duality and its application to holography.

In this paper, we generalize the FZZ-duality by replacing the coset model \eqref{sl2coset} with
\begin{align}
\label{slNcoset}
\frac{\mathfrak{sl}(N+1)_k}{\mathfrak{sl}(N)_k \times \mathfrak{u}(1)} \, .
\end{align}
The central charge is
\begin{align}
c = \frac{k ((N+1)^2 -1)}{k - N - 1} - \frac{k (N^2 -1)}{k- N} - 1 \, .\label{slNcenter}
\end{align}
The symmetry of the coset model \eqref{slNcoset} is given by a one parameter quotient of the W$_\infty$-algebra \cite{Creutzig:2020zaj}.
Recently, W-algebras and especially the W$_\infty$-algebra play important roles in theoretical physics. For instance, subsectors of four dimensional gauge theories are known to be described by Toda field theories with certain W-algebra symmetry \cite{Alday:2009aq,Wyllard:2009hg}.
Moreover, the coset model \eqref{slNcoset} is essentially the one appearing in the higher spin holography \cite{Gaberdiel:2010pz}  or its supersymmetric extension \cite{Creutzig:2011fe}.

This connection to the cosets relevant to higher spin (super-)gravity is our first main motivation. The second motivation is a conformal field theory derivation of dualities and trialities of corner vertex operator algebras (VOAs). Recall that at the corner of interfaces of four dimensional supersymmetric gauge theories with gauge groups $U(N_1), U(N_2)$ and $U(N_3)$ there is a VOA, called the $Y_{N_1, N_2 , N_3}$-algebra \cite{Gaiotto:2017euk}. This vertex algebra is parameterized by a parameter $\psi$ (for the associated CFT it is the inverse of the coupling constant), which is the level of the associated $W$-algebra shifted by the dual Coxeter number. The triality conjecture of \cite{Gaiotto:2017euk} asserts that $Y_{N_1, N_2, N_3}$-algebras are invariant under permutations of the labels where the coupling constant also changes appropriately. Especially the $Y_{N_1, N_2, N_3}$-algebra at coupling $\psi$ coincides conjecturally with the $Y_{N_2, N_1,N_3}$-algebra at coupling $\psi^{-1}$. I.e. if one of the theories is strongly coupled (small $\psi$) the other one is weakly coupled (large $\psi$). 
The triality conjecture is a theorem if one of the labels is zero \cite{Creutzig:2020zaj}.
The simplest example, the $Y_{N, 0, 0}$-algebra is the $W_N$-algebra of $\mathfrak{sl}(N)$ and the isomorphism between $Y_{N, 0, 0}$-algebra and $Y_{0, N, 0}$-algebra is Feigin-Frenkel duality \cite{FF}, while the isomorphism between $Y_{N, 0, 0}$-algebra and $Y_{0, 0, N}$-algebra is the coset realization of the $W_N$-algebra \cite{Arakawa:2018iyk}. The $Y_{N, 1, 0}$-algebra is a coset of the Feigin-Semikhatov algebra of $\mathfrak{sl}(N+1)$ and the isomorphism between 
$Y_{N, 1, 0}$-algebra and $Y_{1, N, 0}$-algebra is the Feigin-Semikhatov duality \cite{Feigin:2004wb}, proven in \cite{Creutzig:2020vbt}.
It is expected that the dualities and trialities of the vertex algebras of protected operators at corners and junctions lift to dualities of spaces of conformal blocks and thus providing a picture that contains the quantum geometric Langlands correspondence \cite{Creutzig:2017uxh,Frenkel:2018dej,Creutzig:2018ltv}. 

It is thus desirable to give conformal field theory derivations of the dualities and trialities of $Y$-algebras. Especially, one would like to not only have an isomorphism of VOAs, but also a duality of full conformal field theories as e.g. a matching of correlation functions. Our higher rank FZZ-duality is precisely this for the case of the $Y_{0,N, N+1}$-algebras. Moreover, the $Y_{N_1, N_2, N_3}$-algebras are conjecturally the same as the $W_{N_1, N_2, N_3}$-algebras of Bershtein, Feigin and Merzon \cite{BFM}, see also \cite{Litvinov:2016mgi,Prochazka:2018tlo}. The latter act on the cohomology of moduli spaces of spiked instantons of Nekrasov \cite{Rapcak:2018nsl}. It is thus also desirable to get a conformal field theory perspective on the conjectural relation between $Y_{N_1, N_2, N_3}$-algebras and $W_{N_1, N_2, N_3}$-algebras. The latter are characterized as intersections of kernels of screening charges acting on a free field theory, i.e. they have a nice description in terms of a free theory with interaction terms. The dual theory that we get is a free field theory with interaction terms and the holomorphic part of our interaction terms are precisely the screening charges of the $W_{N, 0, N+1}$-algebra \cite{Prochazka:2018tlo}, i.e. our formalism naturally relates $Y_{0, N, N+1}$-algebra to $W_{N, 0, N+1}$-algebra.

The $W_{N, 0, N+1}$-algebra is called as W($\mathfrak{sl}(N+1|N)$)-algebra in \cite{BFM,Litvinov:2016mgi} due to its $\mathfrak{sl}(N+1|N)$ structure. 
The W($\mathfrak{sl}(N+1|N)$)-algebra is closely related to the $\mathcal{N}=2$ W$_{N+1}$-algebra, which  is obtained as a Hamiltonian reduction of $\mathfrak{sl}(N+1|N)$ current algebra.
However, note that the two algebras are different.
The $\mathcal{N}=2$ W$_{N+1}$-algebra is identified as the symmetry algebra of $\mathfrak{sl}(N+1|N)$ Toda field theory, and it is believed to be realized by the CP$_N$ Kazama-Suzuki coset model as well \cite{Ito:1990ac,Ito:1991wb}.%
\footnote{This Conjecture is proven in the case $N=2$ \cite{Genra:2019tgw} and by Lemma 7.12 of \cite{Creutzig:2014lsa} both theories have the same type of minimal strong generators.}
From this fact, we also propose a supersymmetric version of our higher rank FZZ-duality, but its detailed analysis is postponed to future work.

In this paper, we explicitly derive the duality with the coset model \eqref{slNcoset} for $N=2,3$ by extending the proof of the original FZZ-duality in \cite{Hikida:2008pe}.
In the proof of FZZ-duality, the $\mathfrak{sl}(2)$ factor of the coset \eqref{sl2coset} described by $\mathfrak{sl}(2)$ Wess-Zumino-Novikov-Witten (WZNW) model  is reduced to Liouville field theory by applying the analysis of \cite{Hikida:2007tq}, which is a path integral derivation of (generalized) Ribault-Teschner relation \cite{Ribault:2005wp,Ribault:2005ms}.
We further utilize the self-duality of Liouville field theory, which is the key point for the nature of strong/weak duality.
There have been several works on generalizations of the reduction method  \cite{Hikida:2007sz,Creutzig:2011qm,Creutzig:2015hla}.
Recently,  big progress was made  in \cite{Creutzig:2020ffn}, and the result enables us to do the current generalizations of FZZ-duality.
In order to directly apply the procedure of \cite{Hikida:2008pe}, we may want to reduce the $\mathfrak{sl}(N+1)$ factor of the coset \eqref{slNcoset} to $\mathfrak{sl}(N+1)$ Toda field theory.  If this is possible, then the self-duality of the Toda field theory can be applied. However, its difficulty has been recognized, see, e.g., \cite{Ribault:2008si,Creutzig:2015hla}.
We can avoid this difficulty by not directly reducing the WZNW model to the Toda field theory but instead using intermediate theories with non-regular type of W-algebra symmetry  \cite{Creutzig:2020ffn}. 
A famous example of non-regular W-algebra is given by Bershadsky-Polyakov (BP-)algebra \cite{Polyakov:1989dm,Bershadsky:1990bg}.
We make use of the intermediate theories in the following way.
The method of \cite{Hikida:2007tq} and its generalizations heavily rely on the free field realizations of algebras.
For non-regular W-algebras, it was noticed that several free field realizations are available \cite{Genra1,Genra2}.
The reduction then becomes possible by replacing its free field realization with a convenient one for the intermediate theories.

The paper is organized as follows.
In the next section, we start by examining the symmetry of the coset model, which is also underlying the duality.
In section \ref{sec:sl3coset}, we study the coset model \eqref{slNcoset} with $N=2$.
We develop a free field realization of  the coset algebra by following \cite{Gerasimov:1989mz,Kuwahara:1989xy}, which is necessary to apply the reduction method of \cite{Hikida:2007tq,Creutzig:2020ffn}.
In section \ref{sec:sl3gfzz}, we derive a generalized FZZ-duality with the coset \eqref{slNcoset} for $N=2$.
In this case, the annoying issue associated with higher rank algebra does not arise and the original reduction method of \cite{Hikida:2007tq} and the self-duality of $\mathfrak{sl}(3)$ Toda field theory can be applied straightforwardly. Massaging the theory obtained in this way, we derive a duality between the coset model \eqref{slNcoset} with $N=2$ and a generalized sine-Liouville theory with an $\mathfrak{sl}(3|2)$ structure.
In section \ref{sec:HRG}, we generalize the analysis  by considering the coset model \eqref{slNcoset} with $N=3$. A free field realization of the coset algebra can be obtained in a way similar to the  $N=2$ case. However, the original reduction method of \cite{Hikida:2007tq} cannot be applied directly. We overcome this difficulty by utilizing the technique of \cite{Creutzig:2020ffn} briefly explained above and derive the dual theory with an $\mathfrak{sl}(4|3)$ structure.
In section \ref{sec:conclusion}, we conclude this paper with an proposal of higher rank FZZ-duality with the coset model \eqref{slNcoset} for generic $N$.
We further discuss open problems and future directions.
In appendix \ref{sec:susy}, we propose an $\mathcal{N}=2$ version of higher rank FZZ-duality based on the results of \cite{Ito:1990ac,Ito:1991wb}. In appendix \ref{sec:BPalgebra}, we summarize useful results on free field realizations of BP-algebra in \cite{Creutzig:2020ffn}.

\section{Symmetry algebra}

As explained in the introduction, a class of VOA  denoted by $Y_{N_1 ,N_2 , N_3}[\psi]$ was introduced in \cite{Gaiotto:2017euk} through brane junctions. There are three types of definitions of the algebra, but  the equivalence of the algebras defined in the three ways is still a conjecture if all labels are non-zero.
The first one is directly obtained from the brane junction picture, and the symmetry algebra of the coset model \eqref{slNcoset} is identified as $Y_{0,N,N+1}[\psi]$ with 
\begin{align}
\psi = - k + N+1  
\end{align}
after decoupling a $\mathfrak{u}(1)$-sector \cite{Gaiotto:2017euk}.
The second one is realized by a truncation of a W$_{1+\infty}[\lambda]$-algebra, which will be explained in the next subsection. The equivalence of these two realizations if one of the three labels is zero has been proven in \cite{Creutzig:2020zaj}.
This realization is used to check whether our formulation of theory possesses desired symmetry algebra.
The third one is given by an intersection of kernels of screening charges in a free field theory, which is actually the definition of $W_{N_1,N_2,N_3}$ mentioned in the introduction. This screening realization has been proven to coincide with the coset definition in a special series of cases \cite{Creutzig:2020vbt}.
The algebra $Y_{0,N,N+1}[\psi]$ is dual to  $Y_{N,0,N+1}[\psi^{-1}]$ \cite{Creutzig:2020zaj}, and the conjectural screening charges for a free field realization of the algebra $Y_{N,0,N+1}[\psi^{-1}]$  are provided in subsection \ref{sec:wslN+1N}.

\subsection{Truncation of W$_{\infty}$-algebra}
\label{sec:symmetry}

In this subsection, we realize the coset algebra by
 a one-parameter subquotient of the $W_\infty[\lambda]$-algebra \cite{Creutzig:2020zaj}. 
Since $\mathfrak{sl}(N)$ with level $k$ may be regarded as $\mathfrak{su}(N)$ with level $-k$, we rewrite the coset \eqref{slNcoset} 
via level-rank duality as
\begin{align}
	\frac{\mathfrak{su}(N+1)_{-k}}{\mathfrak{su}(N)_{-k} \times \mathfrak{u}(1)}  \simeq \frac{\mathfrak{su}(-k)_{N} \times \mathfrak{su}(-k)_1}{\mathfrak{su}(-k)_{N+1}} \, .
\end{align}
If $- k = L$ with integer  $L \geq 2$, then the symmetry of the coset is known to be the usual $W_{L}$-algebra.
However, we are interested in the parameter regime of  $- k < - N - 1$ now, thus the symmetry is an ``analytic continuation'' of the algebra. The extended algebra is a W$_\infty$-algebra called as  $W_\infty[\lambda]$ parametrized by  $\lambda$ and a central charge $c$.
It is rigorously constructed by Andrew Linshaw \cite{Linshaw:2017tvv}.
For the coset \eqref{slNcoset}, $\lambda = - k$ and the central charge is \eqref{slNcenter}.  
It is sometimes convenient to combine a $\mathfrak{u}(1)$-sector to form $W_{1 + \infty}[\lambda]$.

The algebra $W_\infty [\lambda]$ has a spin $s$ current $\mathcal{W}^{(s)}$ with $s=2,3,\ldots$ for generic value of $\lambda$.  An ideal forms at $\lambda = L$ with integer $L \geq 2$ and W$_L$-algebra with the truncation of spin as $s=2,3,\ldots L$ can be obtained by dividing the ideal.
The operator product expansions (OPEs) among generators are unique up to the two parameters $\lambda , c$ mentioned above. The algebra is known to have a triality relation \cite{Gaberdiel:2012ku,Linshaw:2017tvv}  and it is useful to introduce three parameters $\lambda_1 ,\lambda_2 , \lambda_3$ satisfying \cite{Prochazka:2014gqa}
\begin{align}
	\frac{1}{\lambda_1}+ \frac{1}{\lambda_2} + \frac{1}{\lambda_3} = 0 \, .
\end{align}
One of them is identified with $\lambda$ in $W_\infty [\lambda]$ and the central charge is related as
\begin{align}
	c = (\lambda_1 - 1) (\lambda_2 - 1) (\lambda_3 - 1) \, .
\end{align}

In order to read off the parameters $\lambda_i$ from the OPEs, it is convenient to focus on the normalization independent combination of coefficients given by
\begin{align}
 \frac{(C^4_{33})^2 C_{44}^0}{(C_{33}^0)^2} \, .
\end{align}
Here $C_{ij}^k$ represents the coefficient of operator product expansion of $\mathcal{W}^{(i)}\mathcal{W}^{(j)} $ in front of $\mathcal{W}^{(k)}$. We further set $\mathcal{W}^{(0)} = \mathbbm{1}$.
The quantity is related to  $\lambda_i$ by
\begin{align}
	 \frac{(C^4_{33})^2 C_{44}^0}{(C_{33}^0)^2}= \frac{144 (c+2) (\lambda_1 -3) (\lambda_2 -3) (\lambda_3 -3)}{c (5 c + 22) (\lambda_1 -2) (\lambda_2 -2) (\lambda_3 -2)} \, . \label{x21}
\end{align}
In the current case, we may set (see, e.g., (2.19) of \cite{Prochazka:2017qum})
\begin{align}
	\lambda_1= - k \, ,  \quad  \lambda_2 = -\frac{k}{k -N -1} \, , \quad \lambda_3 =  \frac{k}{k- N} \, ,
\end{align}
which leads to
\begin{align} \label{x22}
&  \frac{(C^4_{33})^2 C_{44}^0}{(C_{33}^0)^2} = \frac{1}{2 k^2 (5 N+11)-k (N (5 N+39)+22)+17 N (N+1)} \\ &  \times 
\frac{144 (k-1) (k+3) (N+1) (2 k-3 N) (k-N-1) (k-N) (2 k-N) (4 k-3 (N+1))}{(k+1) (k+2) N (k-2 N) (2 k-N-1) (3 k-2 (N+1)) } \, . \nonumber
\end{align}
We will utilize this to check whether theories have expected symmetries in the analysis below.

Above, we have argued that the algebra $W_\infty[\lambda]$ can be truncated to W$_L$ at $\lambda = L$.
More generically, it is known that there is a truncation when $\lambda_j$ with $j=1,2,3$ satisfy
\begin{align}
\frac{N_1}{\lambda_1} + \frac{N_2}{\lambda_2} + \frac{N_3}{\lambda_3} = 1 \label{truncation}
\end{align}
with non-negative integer $N_i$ \cite{Prochazka:2015deb,Prochazka:2017qum}.
The truncated algebra is supposed to be  $Y_{N_1,N_2,N_3}[\psi]$-algebra.
For the symmetry algebra of the coset model \eqref{slNcoset}, the non-negative integers $N_j$  should be set as
\begin{align}
 N_1 = 0 \, ,  \quad N_2 = N \, , \quad N_3 = N+1 \, .
\end{align}

\subsection{Intersection of kernels of screening charges}
\label{sec:wslN+1N}

Screening charges for free field realization of $Y_{N_1,N_2,N_3}[\psi]$-algebra were proposed in \cite{Prochazka:2018tlo} by extending the previous works of \cite{BFM,Litvinov:2016mgi}. 
In this subsection, we explicitly obtain screening charges for $Y_{N,0,N+1}[\psi^{-1}]$-algebra by following their construction.
There are many equivalent ways to express them but here we use the one resemble to those for $\mathfrak{sl}(N+1|N)$ Toda field theory explained in appendix \ref{sec:susy}.

For $Y_{N_1,N_2,N_3}[\psi]$-algebra, we introduce $(N_1 + N_2 +N_3)$ free bosons with labels 
$\phi^{(\kappa)}_j$ $(\kappa =1,2,3, j=1,2,\ldots, N_\kappa)$. 
The OPEs of the free bosons are normalized as
\begin{align}
 \phi^{(\kappa)}_{j} (z) \phi^{(\kappa ' )}_{j '} (0) \sim -  \frac{h_{\kappa}}{h_1 h_2 h_3} \delta^{\kappa , \kappa '} \delta_{j , j'} \ln z \, .
\end{align}
For our purpose, we set $N_1 = N,N_2 = 0,N_3=N+1$ and (see, e.g., (1.2) of \cite{Prochazka:2018tlo})
\begin{align}
h_1 = i  \sqrt{k - N - 1 } \, , \quad h_2 =  \frac{i }{\sqrt{k - N - 1 }} \, , \quad h_3 = - i \frac{k - N}{\sqrt{k - N - 1 }} \, .
\end{align}
The expression of free field realization is determined by the order of free bosons. We choose the one corresponding to the order $\phi^{(3)}_1 \phi^{(1)}_1 \phi^{(3)}_2 \ldots \phi^{(1)}_N \phi^{(3)}_{N+1}$. In this case, the screening charges are given by
\begin{align}
Q_{l} = \oint d w V_{l} (w) 
\end{align}
for $l=1,2,\ldots , 2N$ with the screening operators as
\begin{align}
V_{2 j -1} (w) = e^{ - h_1 \phi^{(3)}_{j} + h_3 \phi^{(1)}_{j}} \, , \quad
V_{2 j} (w) =  e^{ - h_3 \phi^{(1)}_{j} + h_1 \phi^{(3)}_{j+1}} 
\end{align}
for $j=1,2,\ldots , N$.

Among $(2 N +1)$ free bosons, we decouple a linear combination, $\sum_{j=1}^{N+1} \phi^{(3)}_j + \sum_{j=1}^{N}  \phi^{(1)}_j$, which generates a  $\mathfrak{u}(1)$-sector.
For this, we redefine the free fields as
\begin{align}
\begin{aligned}
&\phi_j = - i (\phi^{(3)}_{j} - \phi^{(3)}_{j+1} ) \quad (j=1,2,\ldots , N) \, , \\
&\varphi_j = i \sqrt{\frac{k - N }{k - N-1}}(\phi^{(1)}_{j} - \phi^{(1)}_{j+1}) \quad (j=1,2,\ldots,N-1) \, , \\
&\chi = \sqrt{\frac{N (N+1)}{2 k}} \left( - \frac{h_1}{N+1}\sum_{j=1}^{N+1} \phi^{(3)}_j + \frac{h_3}{N} \sum_{j=1}^N \phi^{(1)}_j \right ) \, .
\end{aligned}
\end{align}
The OPEs among these fields are
\begin{align}
\phi_i (z) \phi_j (0) \sim - G^{(N+1)}_{ij} \ln z \, , \quad
\varphi_i (z) \varphi_j (0) \sim  G^{(N)}_{ij} \ln z \, , \quad
\chi (z) \chi (0) \sim \frac{1}{2}  \ln z \, ,
 \label{slNOPE}
\end{align}
where $G_{ij}^{(N+1)}$ is the Cartan matrix of $\mathfrak{sl}(N+1)$.
We will use its inverse denoted by  $G^{(N+1)ij}$ as well.
With these new fields, the screening operators are given by 
\begin{align}
& V_1 = e^{\phi^1 /b_{(N+1)} - \varphi^1 /b_{(N)} + \sqrt{\frac{2k}{N(N+1)} } \chi } \, , \nonumber \\
& V_2 = e^{(\phi^1 - \phi^2 ) /b_{(N+1)} + \varphi^1 /b_{(N)} - \sqrt{\frac{2k}{N(N+1)}} \chi } \, , \nonumber \\
& V_3 = e^{(\phi^2 - \phi^1)/b _{(N+1)} + (\varphi^1 - \varphi^2)/b _{(N)} + \sqrt{\frac{2k}{N(N+1)} } \chi }  \, , \nonumber \\
& V_4  = e^{(\phi^2 - \phi^3)/b _{(N+1)} + (\varphi^2 - \varphi^1)/b _{(N)} - \sqrt{\frac{2 k}{N(N+1)} }\chi }  \, ,  \label{bosonicint}\\
& \qquad \qquad \qquad \qquad \vdots \nonumber \\
& V_{2N-1} = e^{(\phi^N - \phi^{N-1})/b _{(N+1)} + \varphi^{N-1} /b _{(N)} + \sqrt{\frac{2k}{N(N+1)} } \chi} \, , \nonumber \\
&V_{2N} = e^{ \phi^N /b _{(N+1)} - \varphi^{N-1} /b _{(N)} - \sqrt{\frac{2 k}{N(N+1)} }\chi }  \, ,\nonumber 
\end{align}
where we have set
\begin{align}
b_{(a)} = \frac{1}{\sqrt{k-a}} \label{bi} 
\end{align}
with $a=N,N+1$.
The indices of $\phi_j$ and $\varphi_j$ are raised as
\begin{align}
\phi^i = G^{(N+1)ij} \phi_j \, , \quad 
\varphi^i = G^{(N)ij} \varphi_j \, .
\end{align}
In subsequent sections, we shall see that the free field realization corresponds to the theory dual to the coset model \eqref{slNcoset} for explicit examples.

\section{Coset model}
\label{sec:sl3coset}

In this section, we examine the coset model \eqref{slNcoset} with focusing on the $N=2$ case given by
\begin{align}
\label{sl3coset}
\frac{\mathfrak{sl}(3)_k}{\mathfrak{sl}(2)_k \times \mathfrak{u}(1)}  \, .
\end{align}
In the next subsection, we explain the first order formulation of $\mathfrak{sl}(3)$ WZNW model and the free field realization of $\mathfrak{sl}(3)$ current algebra.
In subsection \ref{sec:freecoset}, we obtain an action describing the coset model \eqref{sl3coset} and a free field realization of coset algebra.
In subsection \ref{sec:spectral}, we explain the spectral flow automorphism of $\mathfrak{sl}(N+1)$ current algebra.

\subsection{Free field realization of WZNW model}
\label{sec:sl3free}

For the proof of FZZ-duality in  \cite{Hikida:2008pe}, the reduction method of \cite{Hikida:2007tq} was utilized, where the analysis relies on a free field realization of $\mathfrak{sl}(2)$ current algebra.
This implies that a free field realization of coset algebra is necessary in order to apply the analysis to the current case. 
In this subsection, we review the method developed in \cite{Gerasimov:1989mz,Kuwahara:1989xy} and apply it to the coset \eqref{sl3coset}.

The symmetry algebra of the coset \eqref{slNcoset} can be constructed from $\mathfrak{sl}(N+1)$ currents by dividing the denominator algebra, $\mathfrak{sl}(N) \times \mathfrak{u}(1)$. The strategy of \cite{Gerasimov:1989mz,Kuwahara:1989xy} is first expressing the $\mathfrak{sl}(N+1)$ currents by free fields and then reducing the field space such as to be orthogonal to the denominator one.
We thus start from a free field realization of $\mathfrak{sl}(3)$ current algebra.
We use the action of $\mathfrak{sl}(3)$ WZNW model corresponding to a free field realization,
\begin{align}
\label{sl3action}
\begin{aligned}
S &= \frac{1}{2 \pi} \int d^2 w \left[ \frac{G^{(3)}_{ij}}{2}  \partial \phi^i \bar \partial \phi ^j + \frac{b_{(3)}}{4} \sqrt{g} \mathcal{R} (\phi^1 + \phi^2) + \sum_{\alpha =1}^3 (\beta_\alpha \bar \partial \gamma_\alpha + \bar \beta_\alpha \partial \bar \gamma_\alpha) \right] \\
& \quad - \frac{1}{2 \pi k} \int d^2 w \left[ e^{b_{(3)} \phi_1} (\beta_1 - \gamma_2 \beta_3) (\bar \beta_1 - \bar \gamma_2 \bar \beta_3 ) + e^{b _{(3)} \phi_2} \beta_2 \bar \beta_2 \right] \, .
\end{aligned}
\end{align}
Here we set
$
b_{(3)} = 1/\sqrt{k-3} 
$
as in \eqref{bi} with $a=3$ and use the Cartan matrix and its inverse expressed as
\begin{align}
 G_{ij} ^{(3)}= \begin{pmatrix}
 2 & -1 \\
 -1 & 2 
 \end{pmatrix} \, , \quad
  G^{(3)ij} = \begin{pmatrix}
 2/3 & 1/3 \\
 1/3 & 2/3 
 \end{pmatrix} \, . \label{sl3Cartan}
\end{align}
The worldsheet metric is denoted as $g_{\mu \nu}$ with $g = \det g_{\mu \nu}$ and the worldsheet curvature as $\mathcal{R}$.
In terms of OPEs, the free fields satisfy
\begin{align}
 \phi_i (z) \phi_j (0) \sim - G_{ij} ^{(3)}\ln |z|^2 \, , \quad
 \gamma_\alpha (z) \beta_{\alpha '} (0) \sim \frac{\delta_{\alpha , \alpha '}}{z} \, .
\end{align}
The generators of the $\mathfrak{sl}(3)$ current algebra  can be written in terms of the free fields as
(see, e.g., \cite{DiFrancesco:1997nk})
\begin{align}
\begin{aligned}
&e^1 = \beta_1 \, , \quad e^2 = \beta_2 - \gamma_1 \beta_3 \, , \quad e^3 = \beta_3 \, , \\
&h^1 = b^{-1} _{(3)}\partial \phi_1 + 2 \gamma_1 \beta_1 - \gamma_2 \beta_2 + \gamma_3 \beta_3 \, , \\
&h^2 = b^{-1} _{(3)} \partial \phi_2 - \gamma_1 \beta_1 + 2 \gamma_2 \beta_2 + \gamma_3 \beta_3 \, .
\end{aligned}
\end{align}
Here and in the following, the normal ordering prescription is assumed for the products of fields.
The other generators $f^1,f^2,f^3$ can be expressed in a similar manner.
The energy-momentum tensor is given by
\begin{align}
T = - \frac{1}{2} G_{ij}^{(3)} \partial \phi^i \partial \phi^j  + b_{(3)} \partial^2 (\phi^1 + \phi^2) - \sum_{\alpha=1}^3 \partial \gamma_\alpha \beta_\alpha \, .
\end{align}

\subsection{Free field realization of coset model}
\label{sec:freecoset}

Following \cite{Gerasimov:1989mz,Kuwahara:1989xy}, we perform a projection of the numerator algebra, $\mathfrak{sl}(3)$, into a subsector orthogonal to the denominator algebra, $\mathfrak{sl}(2) \times \mathfrak{u}(1)$. 
The generators of $\mathfrak{sl}(2)$ and $\mathfrak{u}(1)$ current algebras may be denoted as $\{\hat h,\hat e,\hat f\}$ and $\hat h'$.
The Cartan subalgebras are generated by
\begin{align}
 \hat h = h^1 + h^2  \, , \quad \hat h' = h^1 - h^2  \, .
\end{align}
For the generators of $\mathfrak{sl}(2)$, we choose $\hat e = e^3$, which determines $\hat f$ as well.
We introduce two free bosons by
\begin{align}
\hat h = 2 b_{(2)}^{-1} \partial \hat \varphi + 2 \gamma_3 \beta_3 \, , \quad \hat h' = 2 \sqrt{3k} \partial \hat \chi \, ,
\end{align}
where $b_{(2)} = 1/\sqrt{k-2}$ as in \eqref{bi} with $a=2$.
This implies that the free bosons satisfy
\begin{align}
\partial \hat \varphi (z) \partial \hat \varphi (0) \sim - \frac{1}{2 z^2} \, , \quad
\partial \hat \chi (z) \partial \hat \chi (0) \sim - \frac{1}{2 z^2} \, .
\end{align}
We define the coset theory by considering the subsector orthogonal to $\hat \varphi , \hat \chi , \gamma_3 , \beta_3 , \bar \gamma_3 , \bar \beta_3$.
For the ghost part with $\gamma_3 , \beta_3 , \bar \gamma_3 , \bar \beta_3 $, we simply decouple it from the rest.

Let us check that the free field realization of the coset theory possesses the correct symmetry algebra.
The free fields are $\phi_1 , \phi_2 $ and the ghost systems $(\gamma_i , \beta_i)$ with $i=1,2$.
We require that the symmetry currents do not have non-trivial OPEs with $\hat \varphi,  \hat \chi$ and
 commute with the interaction terms
\begin{align}
S_1 = \int d^2 w e^{b_{(3)} \phi_1} \beta_1 \bar \beta_1 \, , \quad S_2 =  \int d^2 w e^{b _{(3)} \phi_2} \beta_2 \bar \beta_2 \, .
\end{align}
There are no spin-one currents satisfying these conditions.
We find a spin-two current, which is the energy-momentum tensor,
\begin{align}
\begin{aligned}
T &= - \frac{1}{2} G_{ij}^{(3)} \partial \phi^i \partial \phi^j  + b_{(3) } \partial^2 (\phi^1 + \phi^2) - \sum_{i=1}^2 \partial \gamma_i \beta_i \\
&  \quad + \partial  \hat \varphi \partial \hat \varphi  - b_{(2)}\partial ^2 \hat \varphi + \partial \hat \chi \partial \hat \chi \, .
\end{aligned}
\end{align}
Its central charge reproduces \eqref{slNcenter} with $N=2$. In a similar way, we find a spin-three current and a spin-four current. The ratio of OPE coefficients in \eqref{x21} reproduces \eqref{x22} with $N=2$.

We shall examine the correlation functions of the coset theory.
The vertex operators of the $\mathfrak{sl}(3)$ WZNW model are expressed in the form
\begin{align}
V (z) = P(\gamma_\alpha , \bar \gamma_\alpha ) e^{b_{(3)} (j_1 \phi^1 + j_2 \phi^2 ) } \, ,
\end{align}
where $P(\gamma_\alpha )$ is a function of $\gamma_\alpha  , \bar \gamma_\alpha$ with $\alpha = 1,2,3$.
In the coset theory, we consider a subsector orthogonal to $\hat \varphi , \hat \chi $ and decouple $\gamma_3 ,  \beta_3 , \bar \gamma_3 , \bar \beta_3 $.
We then introduce counter parts $\varphi , \chi$ with negative kinetic terms so as to cancel the contributions from $\hat \varphi , \hat \chi $. The action of the coset theory is now
\begin{align}
S &= \frac{1}{2 \pi} \int d^2 w \left[ \frac{G^{(3)}_{ij}}{2}  \partial \phi^i \bar \partial \phi ^j  - \partial \varphi \bar \partial \varphi -  \partial \chi \partial \chi + \frac{1}{4} \sqrt{g} \mathcal{R} \left(b_{(3)}(\phi^1 + \phi^2)-b_{(2)} \varphi \right) \right] \nonumber  \\ 
& \quad + \frac{1}{2 \pi} \int d^2 w \left[  \sum_{i =1}^2 (\beta_i \bar \partial \gamma_i + \bar \beta_i \partial \bar \gamma_i ) - \frac{1}{k} ( e^{b_{(3)} \phi_1} \beta_1  \bar \beta_1  + e^{b_{(3)} \phi_2} \beta_2 \bar \beta_2 ) \right] \, . \label{cosetaction}
\end{align}
We consider vertex operators of the form
\begin{align}
\begin{aligned}
\Psi (z) & =  \gamma_{1} ^{-j_1 - l -m}  \gamma_2 ^{-j_2 - l + m }   \bar \gamma_{1} ^{-j_1 - l -\bar m }  \bar \gamma_2^{-j_2 -  l + \bar m } e^{b_{(3) }(j_1 \phi_1 + j_2 \phi_2 ) }  e^{2 b_{(2)} l \varphi   +  2 \sqrt{\frac{3}{k}} (m \chi_L + \bar m \chi_R)}  \label{mbasis}
\end{aligned}
\end{align}
such that the contributions from $\hat \varphi , \hat  \chi$ are canceled out by those from $ \varphi , \chi$.
Here we have decomposed as $\chi (z , \bar z) = \chi_L  + \chi_R$.

\subsection{Spectral flow}
\label{sec:spectral}

As in \cite{Hikida:2008pe}, we need to include the effects of spectral flow automorphism of $\mathfrak{sl}(N+1)$ current algebra.
We are using the notation such that commutation relations among mode expansions of generators are given by
\begin{align}
\begin{aligned}
&	[h^i_n , h^j _m] = - 2 k n \delta_{i,j} \delta_{n+m,0} \, , \quad
	[h^i_n , e^j_m] = G_{ij}^{(N+1)} e^j_{n+m } \, ,  \\
&	[h^i_n , f^j_m] = - G_{ij}^{(N+1)} f^j _{n + m} \, , \quad
	[e^i_n , f^j_m] = - \delta_{i,j} h^i_{n + m} + k n \delta_{i,j} \delta_{n+m ,0} \, .
\end{aligned}
\end{align}
The same commutation relations are satisfied even after the changes
\begin{align}
	\rho^{\{S_j\}} (e^i_n) = e^i_{n - S_i} \, , \quad
	\rho^{\{S_j\}} (f^i_n) = f^{i}_{n + S_i} \, , \quad
	\rho^{\{S_j\}} (h^i_{n}) = h^i_{n}  + k S_i \delta_{n,0} 
\end{align}
for $i=1,2,\ldots , N$.
The commutation relations among new generators fix the action of $	\rho^{\{S_j\}} $ to the other generators.
From this, we can see that $\mathfrak{sl}(N+1)$ current algebra possesses $N$-parameter family of spectral flow automorphism.

In the rest of this subsection, we focus on the  $N=2$ case and express $\rho^{S_1 , S_2} \equiv \rho^{\{S_j \}}$.
We define a  state $| S_1 , S_2  \rangle$ satisfying
\begin{align}
	\rho^{S_1 , S_2} (e^\alpha _n) |S_1 , S_2 \rangle = 0 \, , \quad
	 	\rho^{S_1 , S_2} (f^\alpha _n) |S_1 , S_2 \rangle = 0 \, , \quad 
	 	\rho^{S_1 , S_2} (h^i _n) |S_1 , S_2 \rangle = 0 
\end{align}
with $\alpha = 1,2,3$, $i=1,2$ and $n=0,1,2,\ldots$.
We may decompose the state in terms of free fields as
\begin{align}
	|S_1 , S_2 \rangle =  	|S_1 , S_2 \rangle_{(\beta,\gamma)} 	\otimes |S_1 , S_2 \rangle _{\phi} \, .
\end{align}
Here we require
\begin{align}
&\beta_{1, n  -S_1}	|S_1 , S_2 \rangle_{(\beta,\gamma)} = \beta_{2,  n  - S_2}	|S_1 , S_2 \rangle_{(\beta,\gamma)} =\beta_{3, n  - S_1 - S_2}	|S_1 , S_2 \rangle_{(\beta,\gamma)} = 0 \, ,  \\
&\gamma_{1, n +S_1}	|S_1 , S_2 \rangle_{(\beta,\gamma)} = \gamma_{2,  n +S_2}	|S_1 , S_2 \rangle_{(\beta,\gamma)} =\gamma_{3, n +S_1 + S_2}	|S_1 , S_2 \rangle_{(\beta,\gamma)} = 0
\end{align}
for $n = 0 , 1 , 2 , \ldots $ and
\begin{align}
	| S_1 ,S_2 \rangle = e^{ ( S_1 \phi^1 (0)  + S_2 \phi^2 (0) ) /b_{(3)} } | 0 \rangle_\phi \, .
\end{align}
We denote the corresponding operator as $v^{S_1 ,S_2} (\xi)$. The insertion of this operator puts a restriction on the domain of integration over $\beta_1 , \beta_2 , \beta_3$ such that they have zeros of order $S_1, S_2 , S_1 + S_2$, respectively. Moreover, it also induces the insertion of $e^{ ( S_1 \phi^1   + S_2 \phi^2  )  /b _{(3)} } (\xi)$.
See \cite{Hikida:2008pe} for more details in the $N=1$ case.
In the following, we assume that $S_1 , S_2 \geq 0$, which may be realized by utilizing the Weyl symmetry of $\mathfrak{sl}(3)$ Lie algebra.

\section{A generalized FZZ-duality}
\label{sec:sl3gfzz}

Now that we have a free field realization of the coset model \eqref{sl3coset}, 
we can apply the method of \cite{Hikida:2007tq,Hikida:2008pe} to derive a duality.
We review the analysis of \cite{Creutzig:2020ffn} on the reduction from $\mathfrak{sl}(3)$ WZNW model to $\mathfrak{sl}(3)$ Toda field theory in the next subsection, and
 we apply the reduction procedure to the coset theory in subsection \ref{sec:redsl3coset}.
We rewrite the correlators such that they can be interpreted as those of the dual theory with an  $\mathfrak{sl}(3|2)$ structure in subsection \ref{sec:sl3dual}. 

\subsection{Reduction from WZNW model}
\label{sec:redsl3}

In \cite{Creutzig:2020ffn}, we have explored correspondences of theories with different W-algebra symmetry. 
In this subsection, we review a result on the reduction from $\mathfrak{sl}(3)$ WZNW model to $\mathfrak{sl}(3)$ Toda field theory.
The relation of correlation functions obtained by the reduction procedure is generically complicated.
However, as shown in  \cite{Creutzig:2020ffn}, a simplification occurs if we consider a restricted setup where vertex operators are of specific from. In order to obtain the free field realization of the coset theory, we decouple a ghost system $(\gamma_3, \beta_3)$, and this is exactly the condition where the simplification occurs.

We examine correlation functions of $\mathfrak{sl}(3)$ WZNW model of the form
\begin{align}
\begin{aligned}
&\left \langle \prod_{\nu=1}^N V_\nu (z_\nu)  v^{S_1 ,S_2} (\xi)\right \rangle  \\
& \quad = \int_{S_j, \xi } \mathcal{D} \phi_1 \mathcal{D} \phi_2 \left[ \prod_{\alpha =1}^3 \mathcal{D}^2 \beta_\alpha \mathcal{D}^2 \gamma_\alpha \right] 
e^{- S} \prod_{\nu =1}^N V_\nu (z_\nu)  e^{ ( S_1 \phi^1  + S_2 \phi^2 )  /b _{(3)}} (\xi) \, . \label{corr}
\end{aligned}
\end{align}
The domain of integration for $\beta_\alpha$ is restricted as described above.
Here the action is \eqref{sl3action} and the vertex operators are of the form
\begin{align}
V_\nu (z_\nu) = |\mu_1 |^{2 j_1^\nu} |\mu_2|^{2 j_2^\nu} e^{\mu_1^\nu \gamma_1 + \mu_2^\nu \gamma_2 - \bar \mu_1^\nu \bar \gamma_1 - \bar  \mu_2^\nu \bar \gamma_2} e^{b _{(3)} (  j_1^\nu \phi_1 + j_2 ^\nu \phi_2) } \, . \label{mubasis}
\end{align}
Since the vertex operators do not depend on $\gamma_3, \bar \gamma_3$, we can integrate out $\gamma_3, \beta_3, \bar \gamma_3 , \bar \beta_3$. Then, we can simply set $\beta_3 = \bar \beta_3 = 0$, and the action reduces to a part appearing in \eqref{cosetaction}.

With this setup, we can integrate out $\gamma_1 , \gamma_2 , \bar \gamma_1 , \bar \gamma_2$.
This results in the constraints
\begin{align}
 \bar \partial \beta_i (w) = -  2 \pi \sum_{\nu=1}^N \mu^\nu_i \delta^{(2)} (w - z_\nu) \, , \quad
 \partial \bar \beta_i (w) =  2 \pi \sum_{\nu=1}^N \bar \mu^\nu_i \delta^{(2)} (w - z_\nu)
\end{align}
with $i=1,2$.
Solving the equations,  we have
\begin{align}
\beta_i (w) = - \sum_{\nu =1}^N\frac{\mu^\nu_i}{w - z_\nu} = - u_ i \frac{(w - \xi)^{S_i}\prod_{p=1}^{N-2 - S_i} (w -y_p^i )}{ \prod_{\nu =1}^{N} (w - z_\nu ) }  = - u_i \mathcal{B}_i (w ; z_\nu , y_p^i) \label{B}
\end{align}
and similarly for $\bar \beta_i$.
Here we have used the fact that $\beta_i$ should have a zero of order $S_i$ at $w = \xi$ and a holomorphic one-form with $N$ poles should have $(N-2)$ zeros on a Riemann sphere. In order for this to be satisfied, we have to assign 
\begin{align}
\sum_{\nu=1}^N \frac{\mu^\nu_i}{(\xi - z_\nu)^n}= 0
\end{align}
for $n=0,1,\ldots , S_i$.

Inserting \eqref{B} into the action \eqref{sl3action} with $\beta_3 = 0$, the coefficients of interaction terms have now coordinate dependence. We remove the coordinate dependence by shifting $\phi_i$ as
\begin{align}
\phi_i + \frac{1}{b_{(3)}} \ln |u_i \mathcal{B}_i|^2 \to \phi_i
\end{align}
with $i=1,2$. The shifts yield several contributions from the kinetic terms of $\phi_i$ among others, and the correlation function \eqref{corr} becomes of the form
\begin{align}
\begin{aligned}
& \left \langle \prod_{\nu=1}^N V_\nu (z_\nu)   v^{S_1 ,S_2} (\xi) \right \rangle \\
& \quad  = |\Theta_N|^2 \prod_{i=1}^2 \prod_{n=0}^{S_i} \delta^{(2)} \left (\sum_{\nu=1}^N \frac{\mu_i^\nu}{(\xi - z_\nu)^n}  \right) 
\left \langle  \prod_{\nu=1}^N \tilde V_\nu (z_\nu) \prod_{p=1}^{N-2 -S_1} \tilde V_b (y_p^1)  \prod_{p=1}^{N-2 -S_2}  \tilde V_b (y_p^2) \right \rangle \, . 
\end{aligned} \label{reduction}
\end{align}
The right hand side is computed with the action of $\mathfrak{sl}(3)$ Toda field theory,
\begin{align}
S = \frac{1}{2\pi} \int d ^2 w  \left[ \frac{G^{(3)}_{ij}}{2} \partial \phi^i \bar \partial \phi^j + \frac{1}{4} \sqrt{g} \mathcal{R} \left( b_{(3)} + b_{(3)}^{-1}\right) (\phi^1 + \phi^2) + \frac{1}{k } ( e^{b_{(3)} \phi_1} + e^{b_{(3)} \phi_2} )\right] \, .
\end{align}
Later we use the self-duality of $\mathfrak{sl}(3)$ Toda field theory under the exchange of $b_{(3)} \leftrightarrow b_{(3)}^{-1}$.
The modified vertex operators are 
\begin{align}
	\tilde V_\nu (z_\nu) = e^{b_{(3)} (j_1 ^\nu + b_{(3)}^{-2} ) \phi_1 + b_{(3)} ( j_2^\nu + b_{(3)}^{-2} ) \phi_2 }  \, , \quad
	\tilde V_b (y_p^i) = e^{- \phi^i/b _{(3)}} \, .
\end{align}
The pre-factor is computed as
\begin{align}
\begin{aligned}
\Theta_N &= u_1^{2 - \frac{2 S_1 + S_2}{3 b^2_{(3)}}} u_2^{ 2 - \frac{S_1 + 2 S_2}{3 b^2_{(3)}}} \\
& \quad \times
\left[ \prod_{\nu < \nu '}  z_{ \nu  \nu '}\right] ^{ \frac{2}{ b^{2}_{(3)}}}  \left[ \prod_{\nu ,p , i} (z_\nu - y^i_p) \right]^{- \frac{1}{b^{2}_{(3)}} }
 \left[ \prod_{p,p'} (y_p^1 - y_{p'}^2) \right]^{\frac{1}{3 b^{2}_{(3)}}} \left[  \prod_{p < p' , i} y^i_{ p p '} \right] ^{\frac{2}{3  b^{2}_{(3)}}} \, ,
\end{aligned}
\end{align}
where $z_{\nu \nu '} = z_\nu - z_{\nu '}$ and $y^i_{p p'} = y^i_p - y^i_{p'}$.
See \cite{Hikida:2007tq,Hikida:2008pe} for more details on the reduction procedure in the case of $\mathfrak{sl}(2)$ WZNW model.

\subsection{Reduction from coset theory}
\label{sec:redsl3coset}

We apply the reduction relation from $\mathfrak{sl}(3)$ WZNW model to the case with the coset theory by following the analysis of \cite{Hikida:2008pe}. In order to do so, we need to add free bosons $\varphi,\chi$ and use the vertex operators of the form \eqref{mbasis}. It is convenient to use the Mellin transforms of  \eqref{mubasis}, which are given by
\begin{align}
	\label{Phimu}
\Phi_\nu (z_\nu) = \int \frac{d^2 \mu_1^\nu d^2 \mu_2^\nu}{|\mu_1^\nu|^2 |\mu_2^\nu|^2} (\mu_1^\nu)^{l^\nu+m^\nu}(\bar \mu_1^\nu)^{ l^\nu+\bar m^\nu } (\mu_2^\nu)^{ l^\nu - m^\nu } ( \bar \mu_2^\nu) ^{ l^\nu - \bar m^\nu } V_\nu (z_\nu) \, .
\end{align}
The coset vertex operators are then expressed as
\begin{align}
\Psi_\nu (z_\nu) = \Phi_\nu (z_\nu) V^{\varphi,\chi}_{l^\nu ; m^\nu,\bar m^\nu} (z_\nu) \, , \quad 
V^{\varphi,\chi}_{l^\nu ; m^\nu,\bar m^\nu} (z_\nu) = e^{2 b_{(2)} l^\nu \varphi + 2  \sqrt{\frac{3}{k} } (m ^\nu \chi_L + \bar m ^\nu \chi_R)} \, .
\end{align}
In the coset theory, we may insert an identity operator
\begin{align}
\mathbbm{1} = v^{S_1 ,S_2}(\xi ) e^{- (S_1 + S_2)  \varphi / b_{(2)} } (\xi) e^{- \sqrt{\frac{k}{3}} (S_1 - S_2) (\chi_L + \chi_R)} (\xi) \, ,  \label{sl3id}
\end{align}
which uses the spectral flow operator $v^{S_1 ,S_2}(\xi )$. Precisely speaking, we denote $v^{S_1 ,S_2}(\xi )$ here as the one without the effect on $\beta_3$.

Here we perform the reduction procedure in the previous subsection, where the parameters $\mu^\nu_i , \bar \mu^\nu_i$ are replaced by $y_p^i, \bar y_p^i$ with $i = 1,2$.
The Jacobian due to the change of variables is \cite{Ribault:2005wp,Ribault:2005ms,Hikida:2008pe}
\begin{align}
\begin{aligned}
& \prod_{\nu=1}^N \frac{d^2 \mu_i^\nu}{|\mu_i^\nu|^2}  \prod_{n=0}^{S_i} \delta^{(2)} \left (\sum_{\nu=1}^N \frac{\mu_i^\nu}{(\xi - z_\nu)^n } \right ) \\
& \quad  = \frac{\prod_{\nu < \nu ' }^N  |z_{\nu \nu '}|^2 \prod_{p < p'}^{N-2 - S_i} | y^{i}_{p p'} |^2}{\prod_{\nu=1}^N \prod_{p=1}^{N-2-S_i} |z_\nu - y_p^i|^2} \frac{d^2 u_i}{|u_i|^{4 + 2 S_i}} \prod_{p=1} ^{N-2 - S_i}d^2 y_p^i
\end{aligned}
\end{align}
for each $i=1,2$.
Because of the transformation of \eqref{Phimu}, the pre-factors with  $u_i \mathcal{B}_i, \bar u_i \bar{\mathcal{B}}_i$ remain.
In order to remove them, we further shift $\varphi , \chi$  as
\begin{align}
\begin{aligned}
&\varphi + \frac{1}{2 b_{(2)}} \left[ \ln |u_1 \mathcal{B}_1| ^2 + \ln |u_2 \mathcal{B}_2| ^2 \right] \to \varphi \, , \\
&\chi_L + \frac{1}{2}\sqrt{\frac{k}{3}} \left[ \ln (u_1 \mathcal{B}_1 )-  \ln (u_2 \mathcal{B}_2 ) \right] \to \chi_L \, , \\
&\chi_R +  \frac{1}{2}\sqrt{\frac{k}{3}}  \left[ \ln ( \bar u_1 \bar {\mathcal{B}}_1 ) -  \ln (\bar  u_2 \bar {\mathcal{B}}_2 ) \right] \to \chi_R \, .
\end{aligned}
\label{shifts2}
\end{align}
The coset correlation function is now written as
\begin{align} \label{cosetcorr}
&\left \langle \prod_{\nu=1}^N \Psi_\nu (z_\nu) \right \rangle = \int   \prod_{i=1}^2 \frac{ \prod_{p=1}^{N-2-S_i} d^2 y_p^i }{(N-2 - S_i)!} \\
& \times
\left \langle  \prod_{\nu=1}^N  \tilde V_\nu (z_\nu) V^{\varphi , \chi}_{l^\nu- b_{(2)}^{-1} ; m^\nu,\bar m^\nu } (z_\nu) \prod_{p=1}^{N-2-S_1} \tilde V_b (y_p^1) V^{\varphi,\chi}_{\frac{1}{2 b_{(2)}^2} ;  \frac{k}{6} ,  \frac{k}{6}} (y_p^1) \prod_{p=1}^{N-2-S_2} \tilde V_b (y_p^2) V^{\varphi,\chi}_{\frac{1}{2 b_{(2)}^2} ;  -\frac{k}{6} , - \frac{k}{6}} (y_p^2) \right \rangle  \nonumber \, . 
\end{align}
The right hand side is evaluated by the action
\begin{align}
S &= \frac{1}{2 \pi} \int d^2 w \left[ \frac{G^{(3)}_{ij}}{2}  \partial \phi^i \bar \partial \phi ^j  - \partial \varphi \bar \partial \varphi -  \partial \chi \partial \chi + \frac{1}{4} \sqrt{g} \mathcal{R} \left(Q_\phi (\phi^1 + \phi^2) - Q_\varphi \varphi \right) \right] \nonumber  \\ 
& \quad +  \frac{1}{k} \int d^2 w \left[  e^{b_{(3)} \phi_1  }   + e^{ b_{(3)} \phi_2  }  \right] \, .  \label{reducedaction}
\end{align}
The background charges are
\begin{align}
Q_\phi = b_{(3)} + \frac{1}{b_{(3)}} \, , \quad Q_\varphi = b_{(2)} + \frac{2}{b_{(2)}} \, .
\end{align}
The division by $(N-2-S_i)!$ comes from $(N-2-S_i)!$-fold map between $\mu_i^\nu$ and $y_p^i$, see \cite{Hikida:2008pe} as well.
Moreover, note that the pre-factor $\Theta_N$  is canceled with similar factors generated due to the shifts \eqref{shifts2}.

\subsection{Dual theory}
\label{sec:sl3dual}

In the previous subsection, we have expressed the $N$-point function of the coset model \eqref{sl3coset} as in the right hand side of \eqref{cosetcorr}.
In this subsection, we rewrite the correlation function such as to be the $N$-point function of the  theory dual to the coset model \eqref{sl3coset} by following \cite{Hikida:2008pe}.

Firstly, we utilize the self-duality of $\mathfrak{sl}(3)$ Toda field theory, which is an essential point to obtain a strong/weak duality.
Instead of \eqref{reducedaction}, we use%
\footnote{The coefficient $\lambda$ in front of the interaction terms can be set arbitrary by shifting  $\phi_i$. 
This also shifts the overall normalization of vertex operators, which are neglected thorough this paper. Other coefficients in interaction terms appearing below will be set in similar ways.}
\begin{align}
S &= \frac{1}{2 \pi} \int d^2 w \left[ \frac{G^{(3)}_{ij}}{2}  \partial \phi^i \bar \partial \phi ^j  - \partial \varphi \bar \partial \varphi -  \partial \chi \partial \chi + \frac{1}{4} \sqrt{g} \mathcal{R} \left(Q_\phi (\phi^1 + \phi^2) - Q_\varphi \varphi \right) \right] \nonumber  \\ 
& \quad +  \lambda \int d^2 w \left[  e^{\phi_1 / b_{(3)} }   + e^{ \phi_2 /  b_{(3)}  }  \right] \, ,
\end{align}
where $b_{(3)}$ in the interaction terms is replaced by $1/b_{(3)}$.

Secondly, we treat the vertex operators inserted at $w= y^i_p$ as interaction terms.
Namely, we express the correlation function as
\begin{align}
&\left \langle \prod_{\nu=1}^N \Psi_\nu (z_\nu) \right \rangle =\left \langle  \prod_{\nu=1}^N  \mathcal{V}_\nu (z_\nu)  \right \rangle  \, , \quad  \mathcal{V}_\nu (z_\nu) =  \tilde V_\nu (z_\nu) V^{\varphi , \chi}_{l^\nu- b_{(2)}^{-2} ; m^\nu,\bar m^\nu } (z_\nu) \, . \label{newrel}
\end{align}
The action for the right hand side is now
\begin{align}
\begin{aligned}
S &= \frac{1}{2 \pi} \int d^2 w \left[ \frac{G^{(3)}_{ij}}{2}  \partial \phi^i \bar \partial \phi ^j  - \partial \varphi \bar \partial \varphi -  \partial \chi \partial \chi + \frac{1}{4} \sqrt{g} \mathcal{R} \left(Q_\phi (\phi^1 + \phi^2) - Q_\varphi \varphi \right) \right]  \\ 
& \quad + \lambda \int d^2 w \sum_{i=1}^4 V_i
\end{aligned}
\label{newaction}
\end{align}
with four interaction terms
\begin{align}
\begin{aligned}
&V_1 = e^{\phi_1 / b_{(3)}}  \, , \quad
V_2 = e^{\phi_2 / b_{(3)} } \, , \\
&V_3 = e^{- \phi^1 / b_{(3)} + \varphi / b_{(2)}  + \sqrt{\frac{k}{3}} \chi }  \, , \quad
V_4 = e^{- \phi^2 / b_{(3)}+  \varphi  / b_{(2)} - \sqrt{\frac{k}{3}} \chi }  \, . \\
\end{aligned} \label{Vs0}
\end{align}
In this way, we can relate the $N$-point function of the coset model \eqref{sl3coset} with the $N$-point function of a theory with the action \eqref{newaction}.

Finally, we rewrite the action \eqref{newaction} such that the interaction terms \eqref{Vs0} become the same as the screening charges in \eqref{bosonicint}.
For this, we change the  fields  as
\begin{align}
\begin{aligned}
& (k-3) \phi_1 + (k-2)\phi_2 - \frac{2  }{b_{(2)} b_{(3)}} \varphi  \to \phi_1 \, ,  \\
& (k-2) \phi_1 + (k-3)\phi_2 -  \frac{2  }{b _{(2)}b _{(3)}} \varphi \to \phi_2 \, ,  \\
& \frac{1}{b _{(2)} b _{(3)}} (\phi_1 + \phi_2 ) - (2 k -5)  \varphi \to \varphi \, . \\
\end{aligned}
\end{align}
Under these changes, the action \eqref{newaction} becomes
\begin{align}
\begin{aligned}
S &= \frac{1}{2 \pi} \int d^2 w \left[ \frac{G^{(3)}_{ij}}{2}  \partial \phi^i \bar \partial \phi ^j  - \partial \varphi \bar \partial  \varphi -  \partial \chi \partial  \chi + \frac{1}{4} \sqrt{g} \mathcal{R} \left(b_{(3)} (\phi^1 +  \phi^2) -  b_{(2)} \varphi \right) \right]\\
& \quad +  \lambda \int d^2 w \sum_{i=1}^4 V_i 
\end{aligned}
\end{align}
with interaction terms
\begin{align}
\begin{aligned}
&V_1 = e^{((k -3) \phi_1 + (k -2) \phi_2 -  \frac{2}{ b_{(3)} b _{(2)}}  \varphi ) / b_{(3)} } \, , \quad
V_2 = e^{((k -2) \phi_1 + (k -3) \phi_2 -  \frac{2}{ b _{(3)} b _{(2)}} \varphi )/ b_{(3)}  } \, , \\
&V_3 =e^{ \phi^1  / b _{(3)} -  \varphi / b_{(2)} +  \sqrt{\frac{k}{3}} \chi} \, , \quad
V_4 =  e^{ \phi^2 / b _{(3)} -  \varphi / b_{(2)} -  \sqrt{\frac{k}{3}}  \chi} \, . \\
\end{aligned} \label{Vs}
\end{align}

They are still different from the screening charges in \eqref{bosonicint}.
For the original FZZ-duality, reflection relations are applied to interaction terms in order to obtain the sine-Liouville theory with \eqref{sineLiouville}.
We would like to act reflections also in the current case.
Let us focus on a combination of interaction terms
\begin{align}
V_3 V_4 \sim  e^{ 2  \phi / b_{(3)}  -  2   \varphi / b_{(2)} } \equiv e^{ - 2 i \phi '}  \, .  \label{int}
\end{align}
Here we set
\begin{align}
\begin{aligned}
&\phi = \frac12 ( \phi_1 +  \phi_2) \, , \quad \phi_\perp = \frac{1}{2 \sqrt{3}} ( \phi_1 - \phi_2) \, ,\\
&\phi ' = \frac{i}{b_{(3)}} \phi - \frac{i}{b_{(2)}} \varphi \, , \quad \varphi ' = \frac{i}{b_{(2)}} \phi - \frac{i}{b_{(3)}} \varphi \, .
\end{aligned}
\end{align}
The background charges for $\phi '$ and $\varphi '$ are
\begin{align}
Q_{\phi '} = i \, , \quad Q_{\varphi '} =  2 i \frac{b_{(3)}}{b_{(2)}} - i  \frac{b_{(2)}}{b_{(3)}} \, .
\end{align} 
The vertex operator of the form $V_\alpha = e^{2 \alpha \phi '}$ satisfies a reflection relation due to the interaction term \eqref{int} as $V_\alpha = R(\alpha) V_{i - \alpha}$. Using the reflection relation, the vertex operator $\mathcal{V}_\nu = \tilde V_\nu V^{\varphi , \chi}_{l_\nu - b_{(2)}^{-2} ;  m_\nu , \bar m_\nu}$
in \eqref{newrel} can be replaced by%
\footnote{If we use the reflection relation, then we need to insert the factor $R(\alpha)$. In this paper, we neglect it since it can be removed by changing the overall factor. }
\begin{align}
\mathcal{V}_{\nu} (z_\nu)= 
e^{b _{(3)} ( j_2^\nu  \phi_1 + j_1 ^\nu  \phi_2 ) } e^{ 2  b_{(2)} l^\nu  \varphi + 2 \sqrt{\frac{3}{k} } (m ^\nu \chi_L + \bar m ^\nu  \chi_R)} \, .
\end{align}
In a similar way,
 the interactions $V_1,V_2$ in \eqref{Vs} are replaced by
	$V_1 = e^{-\phi_1 / b_{(3)} + 2 \varphi / b_{(2)} } ,V_2  = e^{ -  \phi_2 /  b_{(3)} + 2 \varphi /b_{(2)} } $
with the reflection relation.
Further acting the reflections by $V_3$ and $V_4$, respectively, the interaction terms become
\begin{align}
\begin{aligned}
&\tilde V_1 = V_3 = e^{ \phi^1  / b _{(3)} -  \varphi / b_{(2)} +  \sqrt{\frac{k}{3}}  \chi} \, , \\
&\tilde V_2 = V_2  = e^{ (  \phi^1 - \phi^2 )  / b_{(3)}+  \varphi  / b_{(2)}-  \sqrt{\frac{k}{3}} \chi} \, , \\
&\tilde V_3 = V_1 = e^{ (  \phi^2 - \phi^1 ) / b_{(3)} +   \varphi / b_{(2)}+  \sqrt{\frac{k}{3}} \chi} \, , \\
&\tilde V_4 = V_4 = e^{  \phi^2 / b _{(3)} -  \varphi / b_{(2)} -  \sqrt{\frac{k}{3}} \chi} \, .
\end{aligned}
\end{align}
They are the same as the screening operators in \eqref{bosonicint} with $N=2$ as desired.%
\footnote{It might be convenient to redefine $\tilde X = \tilde X_L - \tilde X_R = - i ( \chi_L + \chi_R ) $  in order to compare with the interaction terms of sine-Liouville theory in \eqref{sineLiouville}. 
Here we choose to use $\chi_{L} , \chi_R$ to make the $\mathfrak{sl}(N+1|N)$ structure clearer.}

We can explicitly check that the dual theory has the same symmetry algebra as that of the coset model \eqref{sl3coset}.
The symmetry generators of the theory are written by the four fields $( \phi_1 , \phi_2 , \varphi , \chi)$ and their OPEs with $\tilde V_i$ should be given by total derivatives.
There are no spin-one currents satisfying these conditions.
We find a spin-two current, which is the energy-momentum tensor,
\begin{align}
\begin{aligned}
T &= - \frac{1}{2} G_{ij}^{(3)} \partial \phi^i \partial \phi^j  + b_{(3)} \partial^2 (\phi^1 + \phi^2)  + \partial  \varphi \partial \varphi  - b_{(2)} \partial ^2 \varphi +  \partial \chi \partial \chi \, .
\end{aligned}
\end{align}
Its central charge reproduces \eqref{slNcenter} with $N=2$. In a similar way, we find a spin-three current and a spin-four current. The ratio of OPE coefficients in \eqref{x21} reproduces \eqref{x22} with $N=2$.

\section{A higher rank generalization}
\label{sec:HRG}

In the previous sections, we have examined the coset model \eqref{slNcoset} with $N=2$ and its duality.
In this section, we extend the analysis for the coset \eqref{slNcoset} with $N=3$,
\begin{align}
	\frac{\mathfrak{sl}(4)_k}{\mathfrak{sl}(3)_k \times \mathfrak{u}(1)} \, . \label{sl4coset}
\end{align}
In the $N=2$ case, the reduction procedure itself was rather straightforward.
Namely, we just integrated the ghost sector and performed the shifts of other fields.
It turns out that the situation is more complicate for a higher rank case.
A main point of \cite{Creutzig:2020ffn} was to utilize a map between different free field realizations for same non-regular W-algebra, and the same technique is needed for the current  case as we explain in this section.

\subsection{Free field realization of coset model}
\label{sec:sl4free}

As in the previous case, we start by finding out a free field realization of the coset model.
For this, we first express the $\mathfrak{sl}(4)$ current algebra in terms of free fields and then consider the subsector orthogonal to the denominator algebra of the coset \eqref{sl4coset}.
We use the action of $\mathfrak{sl}(4)$ WZNW model in the first order formulation,
\begin{align}
	\begin{aligned}
		S &= \frac{1}{2 \pi} \int d^2 w \left[ \frac{G_{ij}^{(4)}}{2} \partial \phi^i \bar \partial \phi^j + \frac{b_{(4)}}{4} \sqrt{g} \mathcal{R} \sum_{i=1}^3 \phi^i  + \sum_{\alpha =1} ^6 (\beta_\alpha \bar \partial \gamma_\alpha + \bar \beta_\alpha \partial \bar \gamma_\alpha ) \right] \\
		& \quad - \frac{1}{2 \pi k} \int d^2 w  \left[ e^{b _{(4)} \phi_1} | \beta_1 |^2 + e^{b _{(4)} \phi_2} |\beta_2 - \gamma_1 \beta_4  |^2 + e^{b _{(4)} \phi_3} |\beta_3 - \gamma_2 \beta_5 - \gamma_4 \beta_6|^2 \right] \, ,
	\end{aligned}
\label{sl4action}
\end{align}
see, e.g., (5.2) of \cite{Creutzig:2020ffn}.
Here we set $b_{(4)} = 1/\sqrt{k - 4}$ as in \eqref{bi} with $a=4$ and use
\begin{align}
	G_{ij} ^{(4)}= \begin{pmatrix}
		2 & -1 & 0 \\
		-1 & 2 & -1 \\
		0 & -1 & 2 
		\end{pmatrix} \, , \quad
	G^{(4) ij} = \begin{pmatrix}
		3/4 & 1/2 & 1/4 \\
		1/2 & 1 & 1/2 \\
		1/4 & 1/2 & 3/4 
		\end{pmatrix} \, .
\end{align}
Among the $\mathfrak{sl}(4)$ currents, the Cartan directions are generated by
\begin{align}
\begin{aligned}
&h^1 = b^{-1}_{(4)} \partial \phi_1 + 2 \gamma_1 \beta_1 - \gamma_2 \beta_2 + \gamma_4 \beta_4 - \gamma_5 \beta _5 + \gamma_6 \beta_6 \, , \\
&h^2 = b^{-1} _{(4)}\partial \phi_2   - \gamma_1 \beta_1 + 2 \gamma_2 \beta_2 - \gamma_3 \beta_3 + \gamma_4 \beta_4  + \gamma_5 \beta_5 \, , \\
&h^3 = b^{-1}_{(4)} \partial \phi_3  - \gamma_2 \beta_2 + 2 \gamma_3 \beta_3 - \gamma_4 \beta_4 + \gamma_5 \beta_5+ \gamma_6 \beta_6 \, .
\end{aligned}
\end{align}

In order to construct the coset theory, we identify the generators of  $\mathfrak{sl}(3)$ subalgebra with
\begin{align}
\hat e ^1 = \beta_4 -  \gamma_3 \beta_6 \,  ,  \quad \hat e^2= \beta_3 \, , \quad \hat e^3 = \beta_6 \, , \quad
 \hat h^1 = h^1 + h^2 \, , \quad \hat h^2  = h^3 \, .
\end{align}
The other generators $\hat f^1  , \hat f^2 , \hat f^3 $ can be constructed from the above generators.
The generator of $\mathfrak{u}(1)$ subalgebra is 
\begin{align}
\hat {h } {}' = h^1 - 2 h^2 - h^3 \, .
\end{align}
We introduce new fields by
\begin{align}
	\begin{aligned}
&	b^{-1} _{(3)} \partial \hat \varphi_1 = b^{-1}_{(4)} (\partial \phi_1 + \partial \phi_2 )  + \gamma_1 \beta_1 + \gamma_2 \beta_2 \, , \\
&	b ^{-1} _{(3)} \partial \hat \varphi_2 = b^{-1} _{(4)}\partial \phi_3 - \gamma_2 \beta_2 + \gamma_5 \beta_5 \ , \\
& 2 \sqrt{6 k} \partial \hat \chi = b^{-1} _{(4)}(\partial \phi_1 - 2 \partial \phi_2 - \partial \phi_3) + 4 (\gamma_1 \beta_1 - \gamma_2 \beta_2 - \gamma_5 \beta_5 ) \, .
 \end{aligned}
\end{align}
We then consider the subspace orthogonal to $\hat \varphi_1, \hat \varphi_2 , \hat \chi $ and neglect the ghost systems $ (\gamma_\alpha , \beta_\alpha )$ with $\alpha = 3,4,6$.

Introducing new fields $\varphi_1, \varphi_2 , \chi $ to cancel the contributions from $\hat \varphi_1, \hat \varphi_2  , \hat \chi$, the action of the coset theory is given by
\begin{align}
	\begin{aligned}
		S &= \frac{1}{2 \pi} \int d^2 w \left[ \frac{G^{(4)}_{ij}}{2} \partial \phi^i \bar \partial \phi^j - \frac{G^{(3)}_{ij}}{2} \partial \varphi^i \bar \partial \varphi^j  - \partial \chi \bar \partial \chi  \right] \\
		& \quad + \frac{1}{2 \pi } \int d^2 w  \left[ \frac{1}{4} \sqrt{g} \mathcal{R} \left(b _{(4)} \sum_{i=1}^3 \phi^i - b_{(3)} \sum_{i=1}^2 \varphi^i \right) +  \sum_{\alpha =1,2,5}  (\beta_\alpha \bar \partial \gamma_\alpha + \bar \beta_\alpha \partial \bar \gamma_\alpha )\right] \\
	& \quad - \frac{1}{2 \pi k } \int d^2 w  \left[ e^{b _{(4)} \phi_1} | \beta_1 |^2 + e^{b _{(4)} \phi_2} |\beta_2 |^2 + e^{b _{(4)} \phi_3} | \gamma_2 \beta_5 |^2 \right] \, .
	\end{aligned} \label{sl4cosetaction}
\end{align}
Here one may notice that an interaction term depends on $\gamma_2 , \bar \gamma_2$, which makes difficult to integrate $\beta_2 , \bar \beta_2$ out. 
The vertex operators are of the form
\begin{align}
	\begin{aligned}
		\Psi (z) & = \left[ \prod_{\alpha = 1,2,5}  \gamma_\alpha^{r_\alpha } \bar \gamma_\alpha^{\bar r_\alpha } \right] 
			e^{b _{(4)} (j_1 \phi_1 + j_2 \phi_2 + j_3 \phi_3) } e^{b _{(3)} ( l_1 \varphi_ 1 + l_2 \varphi_2 )  +  2 \sqrt{\frac{6}{k }}  (m \chi_L + \bar m \chi_R)}
	\end{aligned} \label{sl4cosetvertex}
\end{align}
with
\begin{align}\label{r1r2r5}
	r_1 = - j_1 - l_1 - m \, , \quad
	r_2 = - j_2 + j_3 - l_1 + l_2 + m\, , \quad
	r_5 = - j_3 - l_2 + m 
\end{align} 
and similarly for $\bar r_1, \bar r_2 , \bar r_5$.

The spectral flow operator $v^{\{S_i\}} (\xi) $ can be constructed as explained in subsection \ref{sec:spectral}.
In the current case, there are three parameters $S_i $ with $i =1,2,3$. 
Here we choose the parameters such that $\beta_\alpha$ with $\alpha =1,2,5$ have zeros of order $S_i$ with $i=1,2,3$, respectively, at $w = \xi$ after the insertion of the operator. Along with them, there are similar requirements for $\beta_\alpha$ $(\alpha =3,4,6)$. 
It induces the insertion of $e^{\sum_{i=1}^3 S_i \phi^i/b_{(4)}} (\xi)$ at the same time.

\subsection{Reduction from coset theory}
\label{sec:sl4redWZNW}

We compute the correlation functions of coset theory with vertex operators of the form \eqref{sl4cosetvertex}.
It is convenient to express them as
\begin{align}
&	\Psi_\nu (z_\nu) = \int \frac{d^2 \mu^\nu_2}{|\mu^\nu_2|^2 } (\mu_2^\nu)^{- r_2^\nu} (\bar \mu_2^\nu)^{- \bar r_2^\nu} 
	\tilde \Psi_\nu (z_\nu) \, ,  \label{psi} \\
&	\tilde \Psi_\nu (z_\nu) =  \Phi_\nu (z_\nu) 	V^{\varphi _1 , \varphi_2 , \chi}_{l_1^\nu ,l_2^\nu ; m^\nu , \bar m^\nu} (z_\nu) \label{tildepsi}
\end{align}
with $r_2^\nu$ as in \eqref{r1r2r5}.
Here we have defined $\Phi_\nu (z_\nu)$ as
\begin{align}
&  \Phi_\nu (z_\nu) =   \int \frac{d^2 \mu^\nu_1 d ^2 \mu^\nu_5 }{|\mu^\nu_1|^2 |\mu_5^\nu|^2 } 
	(\mu_1^\nu)^{l_1^\nu + m^\nu} (\bar \mu_1^\nu)^{l_1^\nu + \bar m^\nu} 	(\mu_5^\nu)^{l_2^\nu - m^\nu} (\bar \mu_5^\nu)^{l_2^\nu - \bar m^\nu } 
	V_\nu (z_\nu) \, , \label{phi} \\
&	V_\nu (z_\nu) = |\mu_1^\nu|^{2 j_1^\nu} |\mu_5^\nu|^{2 j_3^\nu}e^{\sum_{\alpha =1,2,5} (\mu_\alpha ^\nu \gamma_\alpha  - \bar \mu_\alpha ^\nu \bar \gamma_\alpha) } e^{b _{(4)} (j_1^\nu \phi_1 + j_2^\nu \phi_2  + j_3^\nu \phi_3)} (z_\nu) \, ,  \label{sl4vertex}
\end{align}
and $	V^{\varphi_1 , \varphi_2 , \chi}_{l_1^\nu ,l_2^\nu ; m^\nu , \bar m^\nu} (z_\nu)$ as
\begin{align}
	V^{\varphi_1 , \varphi_2 , \chi}_{l_1^\nu ,l_2^\nu ; m^\nu , \bar m^\nu} (z_\nu)= e^{b _{(3)} ( l_1 ^\nu \varphi_ 1 + l_2 ^\nu  \varphi_2 )  +  2 \sqrt{\frac{6}{k }}  (m^\nu \chi_L + \bar m^\nu \chi_R)} \, .
\end{align}
We first focus on the part of $\mathfrak{sl}(4)$ WZNW model.
The vertex operators do not have any dependence on $\gamma_3 , \gamma_4 , \gamma_6$, which means that we can set $\beta_\alpha = 0$ with $\alpha =3,4,6$. We would like to perform the reduction procedure by integrating out  $(\gamma_\alpha , \beta_\alpha )$ with $\alpha =1,2,5$. However, as noticed above, it is difficult to do so for $( \gamma_2 , \beta_2 )$ because of the existence of $\gamma_2$ in an interaction term of \eqref{sl4cosetaction}.
In this subsection, we only consider the correlation functions of $\tilde \Psi_\nu (z_\nu)$ in \eqref{tildepsi}.
We will later perform the integration over $( \gamma_2 , \beta_2 )$.

We thus start from the correlation functions of $\mathfrak{sl}(4)$ WZNW model in the form of
\begin{align}
\begin{aligned}
&\left	\langle \prod_{\nu=1}^N V_\nu (z_\nu)  v^{\{S_j \} } (\xi) \right \rangle \\
&\quad = \int _{S_j , \xi} \left[\prod_{i=1}^3 \mathcal{D} \phi_i \right]
 \left [\prod_{\alpha =1,2,5} \mathcal{D} ^2 \beta_\alpha \mathcal{D}^2 \gamma_\alpha \right ] e^{-S} \prod_{\nu=1}^N V_\nu (z_\nu) e^{\sum_{i=1}^3 S_i \phi^i/b_{(4)}} (\xi)\,  , \label{sl4corr}
\end{aligned}
\end{align}
where $S$ denotes the action of $\mathfrak{sl}(4)$ WZNW model in \eqref{sl4action} with $\beta_\alpha = 0$ $(\alpha =3,4,6)$.
We integrate out $\gamma_i, \beta_i $  $(i=1,5)$ and keep $\gamma_2 , \beta_2$ as it is for a while. 
We also set $S_2 = 0$.
Taking care of the insertion of $v^{\{S_j \}} (\xi)$, we find that 
 $\beta_1, \beta_5$ are replaced by functions as
\begin{align}
\begin{aligned}
& \beta_1 (w) = - \sum_{\nu =1}^N \frac{\mu_1 ^\nu}{w - z_\nu} = - u_1 \frac{(w - \xi)^{S_1}\prod_{p =1}^{N-2 - S_1} (w - y_p^1)}{\prod_{\nu=1}^N (w - z_\nu)} \equiv- u_1 \mathcal{B}_1 (w ; z_\nu , y_p^1) \, ,  \\
& \beta_5 (w) = - \sum_{\nu =1}^N \frac{\mu_5 ^\nu}{w - z_\nu} = - u_3 \frac{(w - \xi)^{S_3}\prod_{p =1}^{N-2 - S_3} (w - y_p^3)}{\prod_{\nu=1}^N (w - z_\nu)} \equiv- u_3 \mathcal{B}_3 (w ; z_\nu , y_p^3) \, .
\end{aligned}
\end{align} 
There are also constraints as
\begin{align}
\begin{aligned}
&\sum_{\nu=1}^N \frac{\mu_1^\nu}{(\xi - z_\nu)^n  } = 0  \quad (n=0,1,\ldots , S_1)\, , \\
&\sum_{\nu=1}^N \frac{\mu_5^\nu}{(\xi - z_\nu)^n  } = 0  \quad (n=0,1,\ldots , S_3) \, .
\end{aligned}
\end{align}

As before, we remove functions in interaction terms by shifting fields as 
\begin{align}
	\phi_1 + \frac{1}{b _{(4)}} \ln |u_1 \mathcal{B}_1| \to \phi_1 \, , \quad 
	\phi_3 + \frac{1}{b _{(4)}} \ln |u_3 \mathcal{B}_3| \to \phi_3 \, .
\end{align}
There are several contributions from the kinetic terms in particular and the correlation function \eqref{sl4corr} becomes
\begin{align}
\begin{aligned}
	\left \langle \prod_{\nu=1}^N V_\nu (z_\nu)   v^{\{S_j \} } (\xi)  \right \rangle 	& = |\Theta_N|^2  \prod_{n=0}^{S_1} \delta^{(2)} \left (\sum_{\nu=1}^N \frac{\mu_1^\nu}{(\xi - z_\nu)^n  } \right )
  \prod_{n=0}^{S_3} \delta^{(2)} \left (\sum_{\nu=1}^N \frac{\mu_5^\nu}{(\xi - z_\nu)^n  } \right )\\
& \quad 	\times 
	\left \langle \prod_{\nu=1}^N \tilde V_\nu (z_\nu) \prod_{p=1}^{N-2 -S_1} \tilde V_b (y^1_p)  \prod_{p=1}^{N-2- S_3}\tilde V_b (y^3_p)  \right \rangle \, . 
\end{aligned}
\end{align}
The right hand side is evaluated with the action
\begin{align}
\begin{aligned}
	S &= \frac{1}{2 \pi} \int d^2 w \left[ \frac{G_{ij}^{(4)}}{2} \partial \phi^i \bar \partial \phi^j + \frac{1}{4} \sqrt{g} \mathcal{R} ((b_{(4)} + b^{-1}_{(4)}) (\phi^1 + \phi^3) + b _{(4)} \phi^2 )   \right]  \\
	& \quad + \frac{1}{2 \pi} \int d^2 w  \left[  \beta_2 \bar \partial \gamma_2 +  \bar \beta_2 \partial \bar \gamma_2 + \frac{1}{k} \left( e^{b _{(4)} \phi_1} - e^{b _{(4)} \phi_2} |\beta_2 |^2 + e^{b _{(4)} \phi_3} |\gamma_2 |^2 \right) \right] \, .
\end{aligned}
\label{sl4newaction}
\end{align}
The vertex operators are modified as
\begin{align}
\tilde V_\nu (z_\nu) = e^{\mu_2^\nu \gamma_2 - \bar \mu_2^\nu \bar \gamma_2}e^{b _{(4)} (j_1^\nu \phi_1 + j_2^\nu \phi_2 + j_3^\nu \phi_3) + (\phi^1 + \phi^3)/b_{(4)}} 
\end{align}
and the inserted vertex operators are
\begin{align}
\tilde V_b (y^1_p) = e^{ - \phi^1/b _{(4)}} \, , \quad \tilde V_b (y^3_p) = e^{ - \phi^3/b_{(4)} } \, . 
\end{align}
The pre-factor is computed as
\begin{align}
& \Theta_N = u_1^{3 - \frac{3 S_1 + S_3}{4 b^2 _{(4)}} } u_3^{ 3 - \frac{ S_1 + 3 S_3}{4 b^2 _{(4)}} } \\
&  \times \left  [ \prod_{\nu < \nu '} z_{\nu \nu '} \right ]  ^{\frac{2}{b^2 _{(4)} }}   \left [  \prod_{p ,\nu} (z_\nu - y^1_p ) (z_\nu - y^3_p ) \right ] ^{- \frac{1}{ b^2 _{(4)} }}   \left [  \prod_{p , p' } (y^1_p - y^3_{p'})\right ] ^{\frac{1}{4 b^2 _{(4)} }}   \left [ \prod_{p < p'} y^1_{p p'} y^3_{p p '} \right ] ^{\frac{3}{4b^2 _{(4)}}}  \, . \nonumber
\end{align}

We then apply the analysis to our coset theory with the action \eqref{sl4cosetaction} and the vertex operators of the form \eqref{tildepsi}. For the coset correlation function, we may insert an identity operator
\begin{align}
	\mathbbm{1} = v^{ \{ S_j \} } (\xi) e^{ - ( S_1  \varphi^1 + S_3 \varphi^2 )/{b_{(3)}} } (\xi ) e^{- (S_1  - S_3) \sqrt{\frac{k}{6}} (\chi_L + \chi_R)} (\xi) \, , \label{sl4id}
\end{align}
where the restriction on the domain of integration is set only for $\beta_1 , \beta_5$.
Applying the reduction produce for the part of $\mathfrak{sl}(4)$ WZNW model, the parameters $\mu_1 ^\nu , \mu_5^\nu$ are mapped to $y_p^1 , y_p^3$. We further shift $\varphi_1 , \varphi_2 , \chi$ as
\begin{align}
& \varphi_1 + \frac{1}{b _{(3)}} \ln |u_1 \mathcal{B}_1|^2 \to \varphi_1 \, , \quad
 \varphi_2 + \frac{1}{b_{(3)}} \ln |u_3 \mathcal{B}_3|^2 \to \varphi_2 \, , \\
& \chi_{L} + \frac{1}{2} \sqrt{\frac{k}{6}} \left[ \ln (u_1 \mathcal{B}_1) - \ln (u_3 \mathcal{B}_3) \right]\to \chi_L \, , \quad
\chi_{R} + \frac{1}{2} \sqrt{\frac{k}{6}} \left[ \ln (\bar u_1 \bar{\mathcal{B}}_1) - \ln(\bar u_3 \bar{\mathcal{B}}_3) \right] \to \chi_R  \nonumber
\end{align}
in order to remove $u_1 \mathcal{B}_1^\nu , u_1 \mathcal{B}_5^\nu$ appearing in \eqref{tildepsi} with \eqref{phi}.
The correlation function is now expressed as
\begin{align}
\left	\langle \prod_{\nu =1}^N \tilde \Psi_\nu (z_\nu) \right \rangle  
= \left \langle \prod_{\nu=1}^N \tilde V_\nu (z_\nu) V^{\varphi_1 , \varphi_2 , \chi}_{l_1 ^\nu -  b_{(3)}^{-2},l_2 ^\nu  - b_{(3)}^{-2} ; m ^\nu  ,\bar  m^\nu } (z_\nu)\right \rangle \, .  \label{intermediate}
\end{align}
The right hand side is evaluated with the action
\begin{align}
	\begin{aligned}
		S &= \frac{1}{2 \pi} \int d^2 w \left[ \frac{G^{(4)}_{ij}}{2} \partial \phi^i \bar \partial \phi^j - \frac{G^{(3)}_{ij}}{2} \partial \varphi^i \bar \partial \varphi^j  - \partial \chi \bar \partial \chi\right] \\
		& \quad + \frac{1}{2 \pi } \int d^2 w  \left[ \frac{1}{4} \sqrt{g} \mathcal{R} \left( \sum_{i=1}^3 Q_{\phi,i}\phi^i - \sum_{i=1}^2 Q_{\varphi,i} \varphi^i  \right) + \beta_2 \bar \partial \gamma_2 + \bar \beta_2 \partial \bar \gamma_2 \right] \\
		& \quad + \frac{1}{2 \pi } \int d^2 w  \left[  \frac{1}{k}\left(e^{b _{(4)} \phi_1}  - e^{b _{(4)} \phi_2} |\beta_2 |^2 + e^{b _{(4)} \phi_3} | \gamma_2 |^2  \right)+ 2 \pi \lambda  (V_1 + V_2 )\right] \, ,	\end{aligned} \label{BPpaction}
\end{align}
where
\begin{align}
	Q_{\phi,1} = Q_{\phi,3} = b _{(4)} + \frac{1}{b _{(4)}} \, , \quad Q_{\phi,2} = b _{(4)} \, , \quad Q_{\varphi,1} =Q_{\varphi,2} = b _{(3)} + \frac{1}{b_{(3)}} 
\end{align}
and 
\begin{align}
	V_1 = e^{- \phi^1 /b _{(4)}+ \varphi^1 /b_{(3)} + \sqrt{\frac{k}{6}} (\chi_L+ \chi_R)} \, , \quad	
	V_2 = e^{- \phi^3 /b _{(4)} + \varphi^2 /b_{(3)} - \sqrt{\frac{k}{6}} (\chi_L+ \chi_R)} \, .
\end{align}
In order to obtain \eqref{intermediate}, we have regarded the vertex operators inserted at $w = y_p^1,y_p^3$ as interaction terms as before.

\subsection{Interpretation as extended BP-theory}
\label{sec:BP}

As mentioned above, we have to deal with $\gamma_2$ in an interaction term of the action \eqref{BPpaction} in order to go furthermore.
We would like to apply the technique developed in \cite{Creutzig:2020ffn}. 
For this, it is convenient to make a change of variables as
\begin{align}
	x =  \phi_1 + \frac{2}{3} \phi_2 + \frac{1}{3} \phi_3 \, , \quad
	x_1 = \phi_2 \, , \quad x_2 = \phi_3 \, . \label{xtophi}
\end{align}
The action is now
\begin{align}
	\begin{aligned}
		S &= \frac{1}{2 \pi} \int d^2 w \left[ \frac{3}{8} \partial x \bar \partial x + 
		 \frac{G^{(3)}_{ij}}{2}   \partial x^i \bar \partial x^j - \frac{G^{(3)}_{ij}}{2} \partial \varphi^i \bar \partial \varphi^j  - \partial \chi \bar \partial \chi \right ] \\
	& \quad + \frac{1}{2 \pi} \int d^2 w \left[	\frac{1}{4} \sqrt{g} \mathcal{R} \left(Q_x  x +  \sum_{i=1}^2 Q_{x,i} x^i - \sum_{i=1}^2 Q_{\varphi,i} \varphi^i  \right) + \beta_2 \bar \partial \gamma_2 + \bar \beta_2 \partial \bar \gamma_2\right] \\
		& \quad + \frac{1}{2 \pi } \int d^2 w  \left[ \frac{1}{k} \left(   -  e^{b _{(4)} x_1} |\beta_2 |^2 + e^{b _{(4)} x_2} | \gamma_2 |^2 \right) + 2 \pi \lambda \sum_{i=0}^2  V_i  \right] \, .
	\end{aligned} \label{BPp2action}
\end{align}
The background charges are
\begin{align}
\begin{aligned}
	&Q_x = \frac{3}{2} b _{(4)} + \frac{1}{b _{(4)}} \, , \quad Q_{x,1} = b _{(4)} \, , \quad Q_{x,2} = b _{(4)} + \frac{1}{b_{(4)}} \, , \\
	&Q_{\varphi,1} = Q_{\varphi,2} = b _{(3)} + \frac{1}{b _{(3)}} \, , 
\end{aligned}
\label{BPBC}
\end{align}
and  the interaction terms are
\begin{align}
\begin{aligned}
&	V_0 = e^{(x - \frac{2}{3} x_1 - \frac{1}{3} x_2)/ b_{(4)} } \, , \\
&	V_1 = e^{- \frac{3}{4} x / b _{(4)} + \varphi^1 /b _{(3)} + \sqrt{\frac{k}{6} } (\chi_L + \chi_R) } \, , \\
&	V_2 = e^{- (\frac{1}{4} x + \frac{1}{3} x_1 + \frac{2}{3} x_2 )/b _{(4)}  + \varphi^2 /b _{(3)} - \sqrt{\frac{k}{6}} (\chi_L + \chi_R)} \, . 
\end{aligned} \label{intvertex}
\end{align}

We would like to evaluate correlation functions of the vertex operators of the form \eqref{psi}.
From the relation of correlation functions \eqref{intermediate} and the change of variables \eqref{xtophi},
the vertex operators are now given by
\begin{align}
	&	\Psi_\nu  ' (z_\nu) = \int \frac{d^2 \mu^\nu_2}{|\mu^\nu_2|^2 } (\mu_2^\nu)^{- r_2^\nu} (\bar \mu_2^\nu)^{- \bar r_2^\nu} 
	\tilde V_\nu (z_\nu) V^{\varphi_1 , \varphi_2 , \chi}_{l_1 ^\nu - b_{(3)}^{-2},l_2 ^\nu  - b_{(3)} ^{- 2} ; m ^\nu  ,\bar  m^\nu } (z_\nu)
\end{align}
with
\begin{align}
	\tilde V_\nu (z_\nu) = e^{\mu_2^\nu \gamma_2 - \bar \mu_2^\nu \bar \gamma_2}e^{(b _{(4)} j _1^\nu + \frac{1}{b _{(4) }}) x + (b_{(4)} (j_2^\nu - \frac{2}{3} j_1^\nu) + \frac{1}{3b_{(4)}} ) x_1 + (b _{(4)} (j_3^\nu - \frac{1}{3} j_1^\nu) + \frac{2}{3b _{(4)} })x_2} \, .
\end{align}
As seen in appendix \ref{sec:BPalgebra}, the first two interaction terms in the action \eqref{BPp2action} correspond to  screening operators of BP-algebra in a free field realization. Following \cite{Creutzig:2020ffn}, we shall remove the $\gamma_2$-dependence of interaction term by exchanging the free field realizations of BP-algebra.

For $V_0$, we have used the self-duality with respect to a Liouville field, which may require explanation.
Let us go back to the action \eqref{BPpaction} and perform the change of variables as
\begin{align}
\phi_1 ' = \frac12 \phi_1 \, , \quad \phi_2 ' = \frac12 \phi_1 + \phi_2 \, , \quad \phi_3 ' = \phi_3 \, .
\end{align}
Then the new field $\phi_1 '$ does not have any non-trivial OPEs with the other new fields and its background charge is $Q = b_{(4)} + b_{(4)}^{-1}$. The term $e^{2 b_{(4)} \phi_1 '}$ can be regarded as the interaction term of Liouville field theory with $\phi_1 '$. Therefore, applying the self-duality of the Liouville field theory, the interaction term can be replaced by $e^{2 \phi_1 ' /b_{(4)}}$. Going back to the original fields $\phi_i$ and applying the change of variables in \eqref{xtophi}, we arrive at the expression with $V_0$ in \eqref{intvertex}.

We shall treat the theory with the action \eqref{BPp2action} as an extension of the theory with BP-algebra symmetry.
The action of the BP-theory is
\begin{align}
	\begin{aligned}
		S &= \frac{1}{2 \pi} \int d^2 w \left[ 
		 \frac{G^{(3)}_{ij}}{2}   \partial x^i \bar \partial x^j + \frac{1}{4} \sqrt{g} \mathcal{R}   \sum_{i=1}^2 Q_{x,i} x^i  \right] \\
& \quad - \lambda  \int d^2 w\left[  e^{b _{(4)} x_1} |\beta_2 |^2  -  e^{b _{(4)} x_2} | \gamma_2 |^2  \right] \, ,
	\end{aligned} \label{BP1}
\end{align}
where $Q_{x,1}, Q_{x,2}$ are given in \eqref{BPBC}.
The interaction terms correspond to screening charges \eqref{1stS} in a free field realization of BP-algebra.
In particular, the spin-one current for the BP-algebra is written in terms of free fields as
\begin{align}
	H = \frac{1}{b_{(4)}} (\partial x^2 - \partial x^1 )   - \gamma _2 \beta _2 \, , \label{H1}
\end{align}
see \eqref{1stH}.
We would like to move to the theory corresponding to the other free field realization of BP-algebra.
From the expression of screening operators in \eqref{2ndS}, the action should be given by
\begin{align}
	\begin{aligned}
		S &= \frac{1}{2 \pi} \int d^2 w \left[ 
		 \frac{G^{(3)}_{ij}}{2}   \partial x^i \bar \partial x^j + \frac{1}{4} \sqrt{g} \mathcal{R}   \sum_{i=1}^2 Q_{x,i} x^i  \right] \\
& \quad - \lambda  \int d^2 w\left[  e^{b _{(4)} x_1} |\beta_2 |^2 -  e^{b _{(4)} x_2}  \right] \, .
	\end{aligned} \label{BP2}
\end{align}
In this case, the spin-one current for the BP-algebra is expressed as
\begin{align}
	H = \frac{1}{b _{(4)}} \partial x^1 + \gamma_2 \beta_2 \, , \label{H2}
\end{align}
see \eqref{2ndH}.

In order to move from one description with \eqref{BP1} to the other one with \eqref{BP2}, we should change the vertex operators such as to behave in the same way under the action of  BP-algebra generators. In the current analysis, we neglect overall factors, so it is enough to have the same charge with respect to the spin-one current $H$. For more details, see appendix \ref{sec:BPalgebra} and \cite{Creutzig:2020ffn}.
For $V_1$ and $V_2$, we can use the same vertex operators since they have the same $H$-charge for the both descriptions. For $V_0$ and $\Psi_\nu '$, we should replace them by
\begin{align}
 &	V_0 ' = \int \frac{d^2 \mu_2}{|\mu_2|^2} |\mu_2|^{- \frac{2}{3b^2_{(4)}}} e^{\mu_2 \gamma_2 - \bar \mu_2 \bar \gamma_2} e^{(x - \frac{2}{3} x_1 - \frac{1}{3} x_2)/ b_{(4)}} \label{V0p}\, , \\
	&	\Psi '' _\nu  (z_\nu) = \int \frac{d^2 \mu^\nu_2}{|\mu^\nu_2|^2 } (\mu_2^\nu)^{- \tilde{r}_2^\nu} (\bar \mu_2^\nu)^{- \bar{\tilde{r}}_2^\nu} 
	\tilde{V}_\nu  (z_\nu) V^{\varphi_1 , \varphi_2 , \chi}_{l_1 ^\nu - b_{(3)}^{-2},l_2 ^\nu  - b_{(3)} ^{-2} ; m ^\nu  ,\bar  m^\nu } (z_\nu)
\end{align}
with
\begin{align}
	\tilde r_2^\nu = l_1 ^\nu - l_2 ^\nu - m ^\nu + j_1^\nu - j_2^\nu \, , \quad 	\bar{\tilde r}_2^\nu = l_1 ^\nu - l_2 ^\nu - \bar m ^\nu + j_1^\nu - j_2^\nu 
\end{align}
in order to have the same $H$-charges after the change of description.

\subsection{Reduction from extended BP-theory}
\label{sec:sl4dual}

Now we can use the action, where the interaction terms do not depend on $\gamma_2$.
Thus, we can integrate $(\gamma_2 , \beta_2 )$ out and apply the reduction procedure.
During the change of description, one of the interaction terms, $V_0$, is replaced by $V_0'$ in \eqref{V0p}, which has now $\gamma_2$-dependence.
 We are implicitly treating the interaction term $V_0 '$ in a perturbative way. Namely, we insert vertex operators at $w = z_\nu$  and integrate over $\mu_2^\nu$, where we have set $\nu=n+1,n+2 , \ldots , n+s $. 
The theory may be regarded as a theory with BP-algebra symmetry but gauged by its $\mathfrak{u}(1)$ subalgebra generated by $H$. For the vertex operators at $w = z_\nu$, the $H$-charge is canceled by the one generated by
\begin{align}
\hat H = \frac{1}{4 b_{(4)}} \partial x -  \frac{1}{b_{(3)}} (\partial \varphi^1 - \partial \varphi^2) + \sqrt{\frac{k}{6}} \partial \chi_L \, .
\end{align}
With the interpretation, an identity operator may be represented as
\begin{align}
	\mathbbm{1} = v^{S_2} (\xi) e^{ - S_2 x /(4 b_{(4)})} (\xi) e^{  S_2 (\varphi^1 - \varphi^2)/b_{(3)}} (\xi) e^{ - S_2 \sqrt{\frac{k}{6}} (\chi_L+ \chi_R) } (\xi) \, . \label{BPid}
\end{align}
Here $v^{S_2}(\xi)$ induces the insertion of $e^{S_2 x^1 /b_{(4)}}(\xi)$ and puts a restriction on the domain of integration over $\beta_2$ such as to have a zero of order $S_2$ at $w = \xi$.%
\footnote{The operator $v^{S_2}$ may be obtained by a twist operator $e^{S_2 \int H}$, and the action of the twist operator can be undone by utilizing $\hat H$ as in \eqref{BPid}.  
It is possible to introduce $v^{\{S_i\}}$ given in subsection \ref{sec:spectral} and the identity operators defined in \eqref{sl3id} and \eqref{sl4id} in a similar manner. See, e.g., \cite{Argurio:2000tb} for the case of $\mathfrak{sl}(2)/\mathfrak{u}(1)$ coset.}

Integration over $\gamma_2 , \beta_2$ leads to the replacement
\begin{align}
	\beta_2 (w) = - \sum_{\nu=1}^{N+s} \frac{\mu_2^\nu}{w - z_\nu} = - u_2 \frac{(w - \xi)^{S_2} \prod_{p=1}^{N+s-2 - S_2} (w - y^2_p)}{\prod_{\nu=1}^{N+s} (w - z_\nu ) }  \equiv - u_2 \mathcal{B} (w; z_\nu,y^2_p) 
\end{align}
subject to the constraints
\begin{align}
\sum_{\nu=1}^{N+s} \frac{\mu_2^\nu}{(\xi - z_\nu)^n} = 0
\end{align}
for $n= 0 ,1, \ldots , S_2$.
We then perform the shifts of fields as
\begin{align}
\begin{aligned}
&	x_1 + \frac{1}{b_{(4)}} \ln |  u_2 \mathcal{B}_2|^2 \to x_1 \, , \quad
	x - \frac{1}{3 b_{(4)}} \ln |  u_2 \mathcal{B}_2|^2 \to x \, , \\
&	\varphi_1 - \frac{1}{b_{(3)}}   \ln |  u_2 \mathcal{B}_2|^2 \to \varphi_1 \, , \quad
	\varphi_2 + \frac{1}{b_{(3)}}  \ln |  u_2 \mathcal{B}_2|^2 \to \varphi_2 \, , \\
&	\chi_L + \frac{1}{2} \sqrt{\frac{k}{6}} \ln (u_2 \mathcal{B}_2) \to \chi_L \, , \quad
	\chi_R + \frac{1}{2} \sqrt{\frac{k}{6}} \ln (\bar u_2 \bar{\mathcal{B}}_2) \to \chi_R \, .
\end{aligned}
\end{align}
The correlation function is now summarized as
\begin{align}
	\left	\langle \prod_{\nu =1}^N \Psi_\nu (z_\nu) \right \rangle  
	= \left \langle \prod_{\nu=1}^N \mathcal{V}_\nu (z_\nu) \right \rangle \, ,
\end{align}
where the action for the right hand side is
\begin{align}
\label{BPaction}
		S &= \frac{1}{2 \pi} \int d^2 w \left[ \frac{3}{8} \partial x \bar \partial x + 
		\frac{G^{(3)}_{ij}}{2}   \partial x^i \bar \partial x^j - \frac{G^{(3)}_{ij}}{2} \partial \varphi^i \bar \partial \varphi^j  - \partial \chi \bar \partial \chi \right ] \\
		& \quad + \frac{1}{2 \pi} \int d^2 w \left[	\frac{1}{4} \sqrt{g} \mathcal{R} \left(Q_x  x +  \sum_{i=1}^2 Q_{x,i} x^i - \sum_{i=1}^2 Q_{\varphi,i} \varphi^i   - Q_\chi \chi \right) + 2 \pi \lambda  \sum_{i=1}^ 6 V_i \right] \, . \nonumber
\end{align}
The background charges are 
\begin{align}
\begin{aligned}
&	Q_x = \frac{3 b_{(4)}}{2}  + \frac{3}{4b_{(4)}} \, , \quad Q_{x,1} = b_{(4)}+ \frac{1}{b_{(4)}}  \, , \quad Q_{x,2} = b_{(4)} + \frac{1}{b_{(4)}} \, , \\
&	Q_{\varphi,1} = b_{(3)} \, , \quad Q_{\varphi,2} = b _{(3)} + \frac{2}{b _{(3)}} \, , \quad Q_\chi = \sqrt{\frac{k}{6}} 
\end{aligned}
\end{align}
and  the interaction terms are
\begin{align}
	\begin{aligned}
		&	V_1 = e^{- \frac{3}{4} x / b _{(4)}+ \varphi^1 /b _{(3)} + \sqrt{\frac{k}{6} } (\chi_L + \chi_R) } \, , \\
		&	V_2 = e^{- (\frac{1}{4} x + \frac{1}{3} x_1 + \frac{2}{3} x_2 )/b _{(4)}  + \varphi^2 /b _{(3)} - \sqrt{\frac{k}{6}} (\chi_L + \chi_R)} \, , \\
		&	V_3 = V_0  ' = e^{\frac{3}{4} x /b _{(4)}+ \varphi^1 /b _{(3)} - \varphi^2 /b _{(3)} - \sqrt{\frac{k}{6}} (\chi_L + \chi_R) }  \, , \\
        &V_4 = e^{x_1/b_{(4)}} \, , \\
        & V_5 = e^{x_2/b _{(4)}}\, , \\
        &V_6 = e^{(\frac{1}{4} x - \frac{2}{3}x_1 - \frac{1}{3} x_2 )/b _{(4)} - \varphi^1 /b _{(3)} + \varphi^2 /b _{(3)}  + \sqrt{\frac{k}{6}} (\chi_L + \chi_R) } \, .
	\end{aligned} \label{intvertexend}
\end{align}
Here we have put $V_0 ' $ back to the action.  For $V_4,V_5$, we have used the self-duality of $\mathfrak{sl}(3)$ Toda field theory, which is an important step to derive a strong/weak duality.
The vertex operators are now of the form
\begin{align}
\begin{aligned}
	\mathcal{V}_\nu (z_\nu)& = e^{(b_{(4)} j _1^\nu + \frac{3}{4b_{(4)}}) x + (b_{(4)} (j_2^\nu - \frac{2}{3} j_1^\nu) + \frac{1}{b_{(4)}} ) x_1 + (b_{(4)} (j_3^\nu - \frac{1}{3} j_1^\nu) + \frac{1}{b _{(4)}})x_2}  \\
& \quad 	\times 
	e^{b _{(3)} ( (l_1 ^\nu - \frac{2}{3 b _{(3)}^2} )\varphi_ 1 +(l_2^\nu - \frac{4}{3 b_{(3)}^2} )\varphi_2 )  +  2 \sqrt{\frac{6}{k }}  ((m^\nu - \frac{k}{12} ) \chi_L + ( \bar m^\nu - \frac{k}{12} )\chi_R)} \, .
\end{aligned}
\end{align}

In order to make the interaction terms to be of  the form \eqref{bosonicint}, we rotate fields and perform reflections as in the $N=2$ case.
We first move from $(x,x_1,x_2)$ to $(\phi_1,\phi_2,\phi_3)$ by \eqref{xtophi} and then perform the rotations of fields as
\begin{align}
 &(k-4) \phi_1 +  (k -4) \phi_2 + (k-3 ) \phi_3  -  \frac{1}{b_{(3)} b_{(4)}}( \varphi_1 + \varphi_2 )   \to  \phi_1  \, , \nonumber\\
 & \left(2 -  \frac{k}{2} \right)\phi_1 + (k-3) \phi_2 + \left(\frac{k}{2} - 2 \right) \phi_3 
+ \frac{2}{3 b_{(3)} b_{(4)}} ( \varphi_1 - \varphi_2)  -\sqrt{\frac{2 k }{3}}  \frac{1}{b_{(4)} } \chi  \to  \phi_2 \, , \nonumber \\
& \left( \frac{3 k}{2} - 5 \right) \phi_1  + \left( \frac{k}{2} - 2 \right)\phi_3 -  \frac{1}{3 b_{(3)} b_{(4)}} (5 \varphi_1 + \varphi_2 )  + \sqrt{\frac{2 k }{3}}  \frac{1}{b_{(4)} } \chi    \to  \phi_3  \, , \\
 & \frac{1}{ b_{(3)} b_{(4)}}  \left ( \frac{1}{2}\phi_1 + 2 \phi_2 + \frac{2}{3} \phi_3   \right) + \left(  1 -\frac{k}{3}  \right)\varphi_1 + \left(6-\frac{5 k}{3}\right) \varphi_2 -\sqrt{\frac{2 k }{3}}  \frac{1}{b_{(3)} }  \chi  \to  \varphi_1 \, , \nonumber \\ 
&\frac{1}{2 b_{(3)} b_{(4)}} ( 3 \phi_1 + \phi_3 )  + \left(6-\frac{5 k}{3}\right) \varphi_1 + \left(1-\frac{k}{3}\right) \varphi_2 + \sqrt{\frac{2 k }{3}}  \frac{1}{b_{(3)} }  \chi \to \varphi_2 \, , \nonumber \\
&-  \sqrt{\frac{k}{6}}\frac{1}{ 4  b_{(4)}} \left(\phi_1  - 2 \phi_2 - \phi_3 \right)  +   \sqrt{\frac{k}{6}} \frac{1}{3 b_{(3)}} ( \varphi_1 - \varphi_2 ) 
   + \left(1-\frac{k}{6}\right) \chi \to \chi \, . \nonumber
\end{align}
The action is now
\begin{align}
\begin{aligned}
		S &= \frac{1}{2 \pi} \int d^2 w \left[ \frac{G^{(4)}_{ij}}{2} \partial \phi^i \bar \partial \phi^j - \frac{G^{(3)}_{ij}}{2} \partial \varphi^i \bar \partial \varphi^j  - \partial \chi \bar \partial \chi\right] \\
		& \quad + \frac{1}{2 \pi} \int d^2 w   \left[  \frac{1}{4} \sqrt{g} \mathcal{R} \left( \sum_{i=1}^3 b_{(4)} \phi^i - \sum_{i=1}^2 b_{(3)} \varphi^i  \right) + 2 \pi \lambda \sum_{i=1}^6 \tilde  V_i \right] 
\end{aligned}
\end{align}
with interaction terms
\begin{align}
	\begin{aligned}
&\tilde V_1 = V_1 = e^{\phi^1 /b_{(4)} - \varphi^1 /b_{(3)} + \sqrt{\frac{k}{6} } (\chi_L+ \chi_R)} \, , \\
&\tilde V_2 = e^{(\phi^1 - \phi^2 ) /b_{(4)} + \varphi^1 /b_{(3)} - \sqrt{\frac{k}{6}} (\chi_L+ \chi_R)} \, , \\
&\tilde V_3 = e^{(\phi^2 - \phi^1)/b _{(4)} + (\varphi^1 - \varphi^2)/b _{(3)} + \sqrt{\frac{k}{6} } (\chi_L+ \chi_R)}  \, , \\
&\tilde V_4 = V_6 = e^{(\phi^2 - \phi^3)/b _{(4)} + (\varphi^2 - \varphi^1)/b _{(3)} - \sqrt{\frac{k}{6} } (\chi_L+ \chi_R)}  \, , \\
&\tilde V_5 = e^{(\phi^3 - \phi^2)/b _{(4)} + \varphi^2 /b _{(3)} + \sqrt{\frac{k}{6} } (\chi_L+ \chi_R)} \, , \\
&\tilde V_6 = V_2 = e^{ \phi^3 /b _{(4)} - \varphi^2 /b _{(3)} - \sqrt{\frac{k}{6} } (\chi_L+ \chi_R)}  \, .
	\end{aligned}
\end{align}
Here $\tilde V_2, \tilde V_3, \tilde V_5$ are obtained from $V_5 , V_4 , V_3$, respectively, by performing certain reflection relations as before.
This form of interaction terms is the same as that in \eqref{bosonicint} as desired.
Similarly, the vertex operators can be put in a nice form  as
\begin{align}
	\mathcal{V}_\nu (z_\nu)& = e^{b _{(4)} (j_3^\nu \phi_1 + j_2^\nu \phi_2 + j_1^\nu \phi_3) + b _{(3)} ( l_2^\nu \varphi_ 1 + l_1 ^\nu \varphi_2 )  +  2 \sqrt{\frac{6}{k }}  (m ^\nu \chi_L + \bar m ^\nu \chi_R)} 
\end{align}
by applying a sequence of reflection relations.

\section{Conclusion and discussions}
\label{sec:conclusion}

In this paper, we generalized the FZZ-duality by replacing \eqref{sl2coset} with a higher rank coset \eqref{slNcoset}. A proof of the original FZZ-duality was given by applying the reduction method of \cite{Hikida:2007tq}.
For the higher rank generalizations, we made use of an extended version of the reduction method recently developed in \cite{Creutzig:2020ffn}.
We first find out an free field realization of the coset algebra by following the analysis of \cite{Gerasimov:1989mz,Kuwahara:1989xy}.
We then applied the reduction method to the coset model for $N=2,3$ and derived the higher rank FZZ-duality between the coset model and the theory with an $\mathfrak{sl}(N+1|N)$ structure. 
For the $N=3$ case, we adopted a technique of  \cite{Creutzig:2020ffn} by exchanging the free field realizations of BP-algebra.
During the process, we applied the self-duality of Toda field theory, which was the key point to derive  strong/weak dualities.

Based on the explicit examples, we would like to propose that the theory dual to the  coset model \eqref{slNcoset} with generic $N$ is given by
\begin{align}
\begin{aligned}
		S &= \frac{1}{2 \pi} \int d^2 w \left[ \frac{G^{(N+1)}_{ij}}{2} \partial \phi^i \bar \partial \phi^j - \frac{G^{(N)}_{ij}}{2} \partial \varphi^i \bar \partial \varphi^j  - \partial \chi \bar \partial \chi\right] \\
		& \quad + \frac{1}{2 \pi}\int d^2 w  \left[  \frac{1}{4} \sqrt{g} \mathcal{R} \left( \sum_{j=1}^{N} b_{(N+1)} \phi^j - \sum_{j=1}^{N-1} b_{(N)} \varphi^j  \right) + 2 \pi \lambda  \sum_{l=1}^{2N}  V_l  \right] \, ,
\end{aligned}
\end{align}
where the interaction terms are as in \eqref{bosonicint}.
We can see that the theory reduces to the one obtained in this paper for $N=2,3$ (and also the sine-Liouville theory for $N=1$ if we use  $\tilde X = \tilde X_L - \tilde X_R = - i ( \chi_L + \chi_R ) $). 
It is an important open problem to give a proof of the higher rank FZZ-dualities.
We expect that a generalized version of the technique utilized for $N=3$ is necessary in order to achieve this.

There are several open problems.
In \cite{Hikida:2008pe}, the correspondence among correlation functions was derived even including all pre-factors.
It might be useful to give a derivation of duality with keeping all pre-factors in the current case as well.
A non-trivial point may arise from the map between the free field realizations of  BP-algebra.
We would like to examine the theory with BP-algebra symmetry furthermore in order to understand 
the map more closely.
In appendix \ref{sec:susy}, we discuss an $\mathcal{N}=2$ supersymmetric version of our duality. We would like to extend the current analysis to the supersymmetric case as mentioned in the introduction.
The duality might be derived by following the analysis of \cite{Hori:2001ax}, where the $\mathcal{N}=2$ supersymmetric version of the original FZZ-duality was proven as a mirror symmetry.
This version of duality was utilized to examine singular Calabi-Yau geometry  \cite{Ooguri:1995wj,Giveon:1999zm},
and its higher rank generalizations may be useful for similar purposes.
The FZZ-duality was extended for higher genus Riemann surfaces in \cite{Hikida:2008pe}, and 
similar extensions would be straightforward even for higher rank cases.
Moreover, the FZZ-duality was generalized for the worldsheet of disk topology in \cite{Creutzig:2010bt}.
However, it seems to be difficult to extend the analysis for higher rank cases, since we do not know much about boundary Toda-like field theories.

Our results can be viewed as a conformal field theoretic version of special cases of duality of $Y_{N_1, N_2, N_3}$-algebras where the first two labels are interchanged. These algebras however enjoy a triality and it is of course desirable to get a conformal field theoretic derivation of the complete triality at least in certain low rank cases. We aim to address this issue in future work. 

An important and quite complicated open question are dualities involving the small and large $\mathcal N = 4$ superconformal algebras.
Note that the small and large $\mathcal N = 4$ superconformal algebras coincide with certain cosets involving $\mathfrak{d}(2, 1; \alpha)$ (and $\mathfrak{psl}(2|2)$) at level one \cite{CFL} and the question is if one can derive $\mathcal N=4$ superconformal field theories from WZNW theories of  $\mathfrak{d}(2, 1; \alpha)$ (and $\mathfrak{psl}(2|2)$) at level one.
This is relevant for the AdS/CFT correspondence which is a
an interesting example of strong/weak duality, namely string theory on strongly curved AdS space corresponds to weakly coupled CFT. Recently, it was argued that ``tensionless'' superstrings on AdS$_3 \times$S$^3 \times$T$^4$ with one unit of NSNS-flux is dual to the undeformed symmetric orbifold of T$^4$, see, e.g., \cite{Gaberdiel:2018rqv,Eberhardt:2018ouy,Giribet:2018ada}.
The match of correlation functions was confirmed recently in \cite{Eberhardt:2019ywk,Eberhardt:2020akk,Hikida:2020kil,Dei:2020zui}, and, in particular, the reduction method of \cite{Hikida:2007tq,Hikida:2008pe} from $\mathfrak{sl}(2)$ WZNW model to Liouville field theory was utilized in \cite{Hikida:2020kil}. The higher rank generalizations of reduction method developed in \cite{Creutzig:2020ffn} and this paper are expected to be useful for investigating the AdS/CFT correspondence involving higher dimensional AdS strings. Tensionless strings are believed to be described by higher spin gravity.
The analysis in this paper would be useful to study CFT dual to higher spin gravity as mentioned in the introduction.
Making use of higher spin holography with $\mathcal{N}=4$ supersymmetry \cite{Gaberdiel:2013vva,Gaberdiel:2014cha} or $\mathcal{N}=3$ supersymmetry \cite{Creutzig:2011fe,Creutzig:2013tja,Creutzig:2014ula}, we would like to examine the relation between superstrings and higher spin gravity.

\subsection*{Acknowledgements}

We are grateful to N.~Genra for useful discussions. 
The work of TC is supported by NSERC Grant Number RES0048511.
The work of YH is supported by JSPS KAKENHI Grant Number 16H02182 and 19H01896.

\appendix

\section{Supersymmetric dualities}
\label{sec:susy}

In this appendix, we propose an $\mathcal{N}=2$ supersymmetric version of our higher rank dualities.
We consider the Kazama-Suzuki model \cite{Kazama:1988uz,Kazama:1988qp} of the form
\begin{align}
 \frac{\mathfrak{sl}(N+1)_k \times \mathfrak{so}(2N)_1}{ \mathfrak{sl}(N)_{k-1} \times \mathfrak{u}(1)} \, ,  \label{KScoset}
\end{align}
which can be regarded as the $\mathcal{N}=2$ supersymmetric extension of the coset model \eqref{slNcoset}.
The symmetry of the coset model is believed to be the $\mathcal{N}=2$ W$_{N+1}$-algebra \cite{Ito:1990ac,Ito:1991wb}.
From this, we propose that  the coset model \eqref{KScoset} is dual to the $\mathfrak{sl}(N+1|N)$ Toda field theory with the same symmetry algebra.
In the following, we provide screening charges for the $\mathcal{N}=2$ W$_{N+1}$-algebra, which correspond to interaction terms of the Toda field theory.

We can choose a purely odd simple root system for $\mathfrak{sl}(N+1|N)$ Lie superalgebra.
We may introduce two orthogonal bases $\epsilon_j$ $(j=1,2,\ldots , N+1)$ and $\delta_j$ $(j=1,2,\ldots ,N)$
satisfying
\begin{align}
\epsilon_i \cdot \epsilon_j = \delta_{i,j} \, , \quad
\delta_i \cdot \delta_j =  - \delta_{i,j} \, .
\end{align}
The odd simple roots are expressed as
\begin{align}
\alpha_{2 j-1} = \epsilon_j - \delta_j \, , \quad \alpha_{2 j} = \delta_j - \epsilon_{j+1}
\end{align}
with $j=1,2,\ldots ,N$.

We introduce free bosons $x_a$ and free fermions $\psi_a$ satisfying
\begin{align}
 x_a (z) x_b (0) \sim - \delta_{a,b} \ln z \, , \quad \psi_a (z) \psi_b (0) \sim \frac{\delta_{a,b}}{z} \, . 
\end{align}
There are background charges for the bosonic fields such that screening charges are of dimension one.
We prepare $x_a,\psi_a$ with $a, b=1,2,\ldots , 2N+1$ but a pair of certain combination decouples at the end.
Screening operators are now expressed as \cite{Ito:1990ac,Ito:1991wb}
\begin{align}
V_l = \alpha_l \cdot \psi e^{\alpha_l \cdot x / b_{(N+1)} }
\end{align}
with $l=1,2,\ldots , 2N$ and $b_{(N+1)}$ given in \eqref{bi} with $a=N+1$.

We may redefine the bosonic fields by
\begin{align}
\begin{aligned}
&\phi_i = (\epsilon_i - \epsilon_{i+1} ) \cdot x  \quad (i=1,2,\ldots , N)\, , \\ 
&\varphi_i = (\delta_i - \delta_{i+1} ) \cdot x \quad (i=1,2,\ldots , N-1)\, , \\
& \chi = \frac{1}{N+1} \sum_{i=1}^{N+1} \epsilon_i \cdot x -  \frac{1}{N} \sum_{i=1}^{N} \delta_i \cdot x \, . 
\end{aligned}
\end{align}
We can check that they are orthogonal to the decoupled mode, 
$\sum_{i=1}^{N+1} \epsilon_i \cdot x - \sum_{i=1}^N \delta_i \cdot x$.
The OPEs among these fields are given by \eqref{slNOPE}.
The screening operators are now expressed as
\begin{align}
&V_1 = \alpha_1 \cdot \psi e^{ (\phi^1 - \varphi^1 + \chi) / b_{(N+1)}} \, , \nonumber \\
&V_2 = \alpha_2 \cdot \psi e^{(\phi^1 - \phi^2 + \varphi^1 - \chi) / b_{(N+1)} } \, , \nonumber \\
& V_3 = \alpha_3 \cdot \psi e^{(\phi^2 - \phi^1 +  \varphi^1 - \varphi^2 +  \chi) / b_{(N+1)} } \, , \nonumber \\
& V_4 = \alpha_4 \cdot \psi e^{(\phi^2 -  \phi^3  + \varphi^2 - \varphi^1 - \chi) / b_{(N+1)} } \, , \label{Itoint}\\
& \qquad \qquad \qquad \vdots \nonumber \\
& V_{2N -1} = \alpha_{2 N -1} \cdot \psi e^{( \phi^{N} - \phi^{N-1} +  \varphi^{N-1} +  \chi) / b_{(N+1)} } \, , \nonumber \\
& V_{2N} = \alpha_{2 N} \cdot \psi e^{( \phi^N -  \varphi^{N-1} -  \chi) / b_{(N+1)} } \, .\nonumber 
\end{align}

\section{Free field realizations of BP-algebra}
\label{sec:BPalgebra}

In this appendix, we summarize the results in section 2 of \cite{Creutzig:2020ffn} on the BP-algebra and its free field realizations. The BP-algebra is generated by a spin-one current $H(z)$, two spin-3/2 bosonic currents $G^\pm (z)$, and a spin-two current $T(z)$, which is the energy momentum tensor. The non-trivial OPEs among them are \cite{Bershadsky:1990bg}
\begin{align}
&T(z) T(0) \sim \frac{\frac{1}{2}c}{z^4} + \frac{2 T(0)}{z^2} + \frac{\partial T (0)}{z} \, , \nonumber \\
&T(z) G^\pm (0) \sim \frac{\frac{3}{2} G^\pm (0)}{z^2} + \frac{\partial G^\pm (0)}{z} \, , \quad
T(z) H(0) \sim \frac{H(0)}{z^2} + \frac{\partial H(0)}{z} \, , \nonumber\\
&H(z) H(0) \sim - \frac{\frac13 (2 k -3)}{z^2} \, , \quad H(z) G^\pm (0) \sim \pm \frac{G^\pm (0)}{z} \, ,\\
&G^+ (z) G^- (0) \sim \frac{(k-1)(2k-3)}{z^3} - \frac{3 (k-1) H(0)}{z^2} \nonumber \\
&\qquad \qquad \qquad \quad + \frac{3 H H (0) + (k-3) T(0) - \frac{3}{2}(k-1) \partial H (0)} {z} \, .  \nonumber
\end{align}
The central charge is
\begin{align}
c = 6 (k-3) + 25 + \frac{24}{k-3} \, .
\end{align}

In order to realize the algebra in terms of free fields, we prepare two free bosons $x_j$ $(j=1,2)$ and a ghost system $( \gamma , \beta )$ satisfying
\begin{align}
x_i (z) x_j (0) \sim - G_{ij}^{(3)} \ln z \, , \quad \gamma (z) \beta (0) \sim \frac{1}{z} \, .
\end{align}
There are two types of screening charges for the same BP-algebra.
A type of screening charges are given by \cite{Bershadsky:1990bg}
\begin{align}
Q_1 = \oint d w e^{b_{(3)} x_1}\beta \, , \quad Q_2 = \oint d w e^{b_{(3)} x_2}\gamma \, . \label{1stS}
\end{align}
The generators should commute with these screening charges, and in particular, the spin-one current is 
\begin{align}
H =  \frac{1}{b_{(3)}} (\partial x^2 - \partial x^1) - \gamma \beta \, . \label{1stH}
\end{align}
The index of $x_j$ is raised by $G^{(3)ij}$, where the expression of $G^{(3)ij}$ can be found in \eqref{sl3Cartan}.
Another type of screening charges are \cite{Genra1,Genra2,Creutzig:2020ffn}
\begin{align}
Q_1 = \oint d w e^{b_{(3)} x_1}\beta \, , \quad Q_2 = \oint d w e^{b_{(3)} x_2} \, . \label{2ndS}
\end{align}
In this case, the spin-one current is obtained as
\begin{align}
H = \frac{1}{b_{(3)}} \partial x^1 + \gamma \beta \, . \label{2ndH}
\end{align}

In the main context, the energy momentum tensor is given by the twisted one
\begin{align}
T_t (z) = T(z) + \frac{1}{2} \partial H(z) \, .
\end{align}
With respect to $T_t(z)$, the conformal dimensions of $G^+(z)$ and $G^-(z)$ become one and two, respectively.
In particular, the generators have integer modded expansions as
\begin{align}
\begin{aligned}
&T_t (z) = \sum_{n \in \mathbb{Z}} \frac{L_n}{z^{n+2} } \, , \quad
H (z) = \sum_{n \in \mathbb{Z}} \frac{H_n}{z^{n+1}} \, , \\
&G^+ (z) = \sum_{n \in \mathbb{Z}} \frac{G^+_n}{z^{n+1}} \, , \quad
G^- (z) = \sum_{n \in \mathbb{Z}} \frac{G^-_n}{z^{n+2}} \, .
\end{aligned}
\end{align}
Primary states with respect to the algebra are defined such as to satisfy the conditions
\begin{align}
L_n |c_2,c_3,m \rangle =  G^\pm_n  |c_2,c_3,m \rangle =  H_n  |c_2,c_3,m \rangle = 0
\end{align}
for $n=1,2,3,\ldots$. The representation of the primary state is labeled by the eigenvalues of the second-order and third-order Casimir operators denoted as $c_2,c_3$.  Here the second-order Casimir operator is given by $L_0$ and the expression of the third-order Casimir operator can be found in \cite{Creutzig:2020ffn}.
The other parameter $m$ is defined by
\begin{align}
H_ 0  |c_2,c_3,m \rangle =  m | c_2 , c_3 , m \rangle \, , 
\end{align}
and in particular
\begin{align}
G_0^\pm  |c_2,c_3,m \rangle \propto  |c_2,c_3,m \pm 1  \rangle \, .
\end{align}

In terms of free fields, the primary states may be expressed by vertex operators as
\begin{align}
  |c_2,c_3,m \rangle \propto \lim_{z \to 0} V_{j_1 , j_2 , \alpha} (z) | 0 \rangle \, , 
\quad V_{j_1 , j_2 , \alpha} (z) = \gamma^\alpha e^{b_{(3)} (j_1 \phi_1 + j_2 \phi_2)} \, .
\end{align}
The parameters $c_2 , c_3$ are independent of both of $\alpha$ and the choice of free field realizations, see \cite{Creutzig:2020ffn} for explicit expressions.
The other parameter $m$ depends on both of $\alpha$ and the choice of free field realizations.
For the free field realization with \eqref{1stS}, the eigenvalue of $H$ is given by
\begin{align}
m = - j_2 + j_1 + \alpha \, .
\end{align}
For the other realization with \eqref{2ndS}, the eigenvalue of $H$ is 
\begin{align}
m = - j_1 - \alpha  \, .
\end{align}
These results imply that we should replace $\alpha \leftrightarrow - 2 j_1 + j_2 - \alpha$ when we exchange the free field realizations.


\begin{thebibliography}{10}

\bibitem{FZZ}
V.~Fateev, A.~Zamolodchikov and A.~Zamolodchikov{,} {unpublished}.

\bibitem{Witten:1991yr}
E.~Witten, \emph{{On string theory and black holes}},
  \href{https://doi.org/10.1103/PhysRevD.44.314}{\emph{Phys. Rev. D} {\bfseries
  44} (1991) 314}.

\bibitem{Kazakov:2000pm}
V.~Kazakov, I.~K. Kostov and D.~Kutasov, \emph{{A matrix model for the
  two-dimensional black hole}},
  \href{https://doi.org/10.1016/S0550-3213(01)00606-X}{\emph{Nucl. Phys. B}
  {\bfseries 622} (2002) 141}
  [\href{https://arxiv.org/abs/hep-th/0101011}{{\ttfamily hep-th/0101011}}].

\bibitem{Creutzig:2020zaj}
T.~Creutzig and A.~R. Linshaw, \emph{{Trialities of $\mathcal{W}$-algebras}},
  \href{https://arxiv.org/abs/2005.10234}{{\ttfamily 2005.10234}}.

\bibitem{Alday:2009aq}
L.~F. Alday, D.~Gaiotto and Y.~Tachikawa, \emph{{Liouville correlation
  functions from four-dimensional gauge theories}},
  \href{https://doi.org/10.1007/s11005-010-0369-5}{\emph{Lett. Math. Phys.}
  {\bfseries 91} (2010) 167} [\href{https://arxiv.org/abs/0906.3219}{{\ttfamily
  0906.3219}}].

\bibitem{Wyllard:2009hg}
N.~Wyllard, \emph{{$A_{N-1}$ conformal Toda field theory correlation functions
  from conformal $\mathcal{N} = 2$ $SU(N)$ quiver gauge theories}},
  \href{https://doi.org/10.1088/1126-6708/2009/11/002}{\emph{JHEP} {\bfseries
  11} (2009) 002} [\href{https://arxiv.org/abs/0907.2189}{{\ttfamily
  0907.2189}}].

\bibitem{Gaberdiel:2010pz}
M.~R. Gaberdiel and R.~Gopakumar, \emph{{An AdS$_3$ dual for minimal model
  CFTs}}, \href{https://doi.org/10.1103/PhysRevD.83.066007}{\emph{Phys. Rev. D}
  {\bfseries 83} (2011) 066007}
  [\href{https://arxiv.org/abs/1011.2986}{{\ttfamily 1011.2986}}].

\bibitem{Creutzig:2011fe}
T.~Creutzig, Y.~Hikida and P.~B. R\o{}nne, \emph{{Higher spin AdS$_3$
  supergravity and its dual CFT}},
  \href{https://doi.org/10.1007/JHEP02(2012)109}{\emph{JHEP} {\bfseries 02}
  (2012) 109} [\href{https://arxiv.org/abs/1111.2139}{{\ttfamily 1111.2139}}].

\bibitem{Gaiotto:2017euk}
D.~Gaiotto and M.~Rap\v{c}\'ak, \emph{{Vertex algebras at the corner}},
  \href{https://doi.org/10.1007/JHEP01(2019)160}{\emph{JHEP} {\bfseries 01}
  (2019) 160} [\href{https://arxiv.org/abs/1703.00982}{{\ttfamily
  1703.00982}}].

\bibitem{FF}
B.~Feigin and E.~Frenkel, \emph{{ Duality in W -algebras }},
  \href{https://doi.org/10.1155/S1073792891000119}{\emph{International
  Mathematics Research Notices} {\bfseries 1991} (1991) 75}
  [\href{https://arxiv.org/abs/https://academic.oup.com/imrn/article-pdf/1991/6/75/6767866/1991-6-75.pdf}{{\ttfamily
  https://academic.oup.com/imrn/article-pdf/1991/6/75/6767866/1991-6-75.pdf}}].

\bibitem{Arakawa:2018iyk}
T.~Arakawa, T.~Creutzig and A.~R. Linshaw, \emph{{W-algebras as coset vertex
  algebras}}, \href{https://doi.org/10.1007/s00222-019-00884-3}{\emph{Invent.
  Math.} {\bfseries 218} (2019) 145}
  [\href{https://arxiv.org/abs/1801.03822}{{\ttfamily 1801.03822}}].

\bibitem{Feigin:2004wb}
B.~Feigin and A.~Semikhatov, \emph{{$\mathcal{W}^{(2)}_n$ algebras}},
  \href{https://doi.org/10.1016/j.nuclphysb.2004.06.056}{\emph{Nucl. Phys. B}
  {\bfseries 698} (2004) 409}
  [\href{https://arxiv.org/abs/math/0401164}{{\ttfamily math/0401164}}].

\bibitem{Creutzig:2020vbt}
T.~Creutzig, N.~Genra and S.~Nakatsuka, \emph{{Duality of subregular
  $\mathcal{W}$-algebras and principal $\mathcal{W}$-superalgebras}},
  \href{https://arxiv.org/abs/2005.10713}{{\ttfamily 2005.10713}}.

\bibitem{Creutzig:2017uxh}
T.~Creutzig and D.~Gaiotto, \emph{{Vertex algebras for S-duality}},
  \href{https://doi.org/10.1007/s00220-020-03870-6}{\emph{Commun. Math. Phys.}
  {\bfseries 379} (2020) 785}
  [\href{https://arxiv.org/abs/1708.00875}{{\ttfamily 1708.00875}}].

\bibitem{Frenkel:2018dej}
E.~Frenkel and D.~Gaiotto, \emph{{Quantum Langlands dualities of boundary
  conditions, D-modules, and conformal blocks}},
  \href{https://doi.org/10.4310/CNTP.2020.v14.n2.a1}{\emph{Commun. Num. Theor.
  Phys.} {\bfseries 14} (2020) 199}
  [\href{https://arxiv.org/abs/1805.00203}{{\ttfamily 1805.00203}}].

\bibitem{Creutzig:2018ltv}
T.~Creutzig, D.~Gaiotto and A.~R. Linshaw, \emph{{S-duality for the Large $N =
  4$ Superconformal Algebra}},
  \href{https://doi.org/10.1007/s00220-019-03673-4}{\emph{Commun. Math. Phys.}
  {\bfseries 374} (2020) 1787}
  [\href{https://arxiv.org/abs/1804.09821}{{\ttfamily 1804.09821}}].

\bibitem{BFM}
M.~Bershtein, B.~Feigin and G.~Merzon, \emph{Plane partitions with a ``pit'':
  generating functions and representation theory},
  \href{https://doi.org/10.1007/s00029-018-0389-z}{\emph{Selecta Math. (N.S.)}
  {\bfseries 24} (2018) 21}.

\bibitem{Litvinov:2016mgi}
A.~Litvinov and L.~Spodyneiko, \emph{{On W algebras commuting with a set of
  screenings}}, \href{https://doi.org/10.1007/JHEP11(2016)138}{\emph{JHEP}
  {\bfseries 11} (2016) 138}
  [\href{https://arxiv.org/abs/1609.06271}{{\ttfamily 1609.06271}}].

\bibitem{Prochazka:2018tlo}
T.~Proch\'azka and M.~Rap\v{c}\'ak, \emph{{$ \mathcal{W} $-algebra modules,
  free fields, and Gukov-Witten defects}},
  \href{https://doi.org/10.1007/JHEP05(2019)159}{\emph{JHEP} {\bfseries 05}
  (2019) 159} [\href{https://arxiv.org/abs/1808.08837}{{\ttfamily
  1808.08837}}].

\bibitem{Rapcak:2018nsl}
M.~Rap\v{c}\'ak, Y.~Soibelman, Y.~Yang and G.~Zhao, \emph{{Cohomological Hall
  algebras, vertex algebras and instantons}},
  \href{https://doi.org/10.1007/s00220-019-03575-5}{\emph{Commun. Math. Phys.}
  {\bfseries 376} (2019) 1803}
  [\href{https://arxiv.org/abs/1810.10402}{{\ttfamily 1810.10402}}].

\bibitem{Ito:1990ac}
K.~Ito, \emph{{Quantum Hamiltonian reduction and $\mathcal{N}=2$ coset
  models}}, \href{https://doi.org/10.1016/0370-2693(91)90136-E}{\emph{Phys.
  Lett. B} {\bfseries 259} (1991) 73}.

\bibitem{Ito:1991wb}
K.~Ito, \emph{{$\mathcal{N}=2$ superconformal CP$_n$ model}},
  \href{https://doi.org/10.1016/0550-3213(92)90347-E}{\emph{Nucl. Phys. B}
  {\bfseries 370} (1992) 123}.

\bibitem{Genra:2019tgw}
N.~Genra and A.~R. Linshaw, \emph{{Ito's conjecture and the coset construction
  for ${\mathcal W}^k(\mathfrak{sl}(3|2))$}}, {\emph{RIMS Kokyuroku Bessatsu}
  (2019) } [\href{https://arxiv.org/abs/1901.02397}{{\ttfamily 1901.02397}}].

\bibitem{Creutzig:2014lsa}
T.~Creutzig and A.~R. Linshaw, \emph{{Cosets of affine vertex algebras inside
  larger structures}},
  \href{https://doi.org/10.1016/j.jalgebra.2018.10.007}{\emph{J. Algebra}
  {\bfseries 517} (2019) 396}
  [\href{https://arxiv.org/abs/1407.8512}{{\ttfamily 1407.8512}}].

\bibitem{Hikida:2008pe}
Y.~Hikida and V.~Schomerus, \emph{{The FZZ-duality conjecture: A proof}},
  \href{https://doi.org/10.1088/1126-6708/2009/03/095}{\emph{JHEP} {\bfseries
  03} (2009) 095} [\href{https://arxiv.org/abs/0805.3931}{{\ttfamily
  0805.3931}}].

\bibitem{Hikida:2007tq}
Y.~Hikida and V.~Schomerus, \emph{{$H_3^+$ WZNW model from Liouville field
  theory}}, \href{https://doi.org/10.1088/1126-6708/2007/10/064}{\emph{JHEP}
  {\bfseries 10} (2007) 064} [\href{https://arxiv.org/abs/0706.1030}{{\ttfamily
  0706.1030}}].

\bibitem{Ribault:2005wp}
S.~Ribault and J.~Teschner, \emph{{$H_3^+$-WZNW correlators from Liouville
  theory}}, \href{https://doi.org/10.1088/1126-6708/2005/06/014}{\emph{JHEP}
  {\bfseries 06} (2005) 014}
  [\href{https://arxiv.org/abs/hep-th/0502048}{{\ttfamily hep-th/0502048}}].

\bibitem{Ribault:2005ms}
S.~Ribault, \emph{{Knizhnik-Zamolodchikov equations and spectral flow in
  AdS$_3$ string theory}},
  \href{https://doi.org/10.1088/1126-6708/2005/09/045}{\emph{JHEP} {\bfseries
  09} (2005) 045} [\href{https://arxiv.org/abs/hep-th/0507114}{{\ttfamily
  hep-th/0507114}}].

\bibitem{Hikida:2007sz}
Y.~Hikida and V.~Schomerus, \emph{{Structure constants of the $OSP(1|2)$ WZNW
  model}}, \href{https://doi.org/10.1088/1126-6708/2007/12/100}{\emph{JHEP}
  {\bfseries 12} (2007) 100} [\href{https://arxiv.org/abs/0711.0338}{{\ttfamily
  0711.0338}}].

\bibitem{Creutzig:2011qm}
T.~Creutzig, Y.~Hikida and P.~B. R\o{}nne, \emph{{Supergroup - extended super
  Liouville correspondence}},
  \href{https://doi.org/10.1007/JHEP06(2011)063}{\emph{JHEP} {\bfseries 06}
  (2011) 063} [\href{https://arxiv.org/abs/1103.5753}{{\ttfamily 1103.5753}}].

\bibitem{Creutzig:2015hla}
T.~Creutzig, Y.~Hikida and P.~B. R\o{}nne, \emph{{Correspondences between WZNW
  models and CFTs with W-algebra symmetry}},
  \href{https://doi.org/10.1007/JHEP02(2016)048}{\emph{JHEP} {\bfseries 02}
  (2016) 048} [\href{https://arxiv.org/abs/1509.07516}{{\ttfamily
  1509.07516}}].

\bibitem{Creutzig:2020ffn}
T.~Creutzig, N.~Genra, Y.~Hikida and T.~Liu, \emph{{Correspondences among CFTs
  with different W-algebra symmetry}},
  \href{https://doi.org/10.1016/j.nuclphysb.2020.115104}{\emph{Nucl. Phys. B}
  {\bfseries 957} (2020) 115104}
  [\href{https://arxiv.org/abs/2002.12587}{{\ttfamily 2002.12587}}].

\bibitem{Ribault:2008si}
S.~Ribault, \emph{{On $sl_3$ Knizhnik-Zamolodchikov equations and $W_3$
  null-vector equations}},
  \href{https://doi.org/10.1088/1126-6708/2009/10/002}{\emph{JHEP} {\bfseries
  10} (2009) 002} [\href{https://arxiv.org/abs/0811.4587}{{\ttfamily
  0811.4587}}].

\bibitem{Polyakov:1989dm}
A.~M. Polyakov, \emph{{Gauge transformations and diffeomorphisms}},
  \href{https://doi.org/10.1142/S0217751X90000386}{\emph{Int. J. Mod. Phys. A}
  {\bfseries 5} (1990) 833}.

\bibitem{Bershadsky:1990bg}
M.~Bershadsky, \emph{{Conformal field theories via Hamiltonian reduction}},
  \href{https://doi.org/10.1007/BF02102729}{\emph{Commun. Math. Phys.}
  {\bfseries 139} (1991) 71}.

\bibitem{Genra1}
N.~Genra, \emph{Screening operators for {$\mathcal{W}$}-algebras},
  \href{https://doi.org/10.1007/s00029-017-0315-9}{\emph{Selecta Math. (N.S.)}
  {\bfseries 23} (2017) 2157}
  [\href{https://arxiv.org/abs/1606.00966}{{\ttfamily 1606.00966}}].

\bibitem{Genra2}
N.~Genra, \emph{Screening operators and parabolic inductions for affine
  $\mathcal{W}$-algebras},  \href{https://arxiv.org/abs/1806.04417}{{\ttfamily
  1806.04417}}.

\bibitem{Gerasimov:1989mz}
A.~Gerasimov, A.~Marshakov and A.~Morozov, \emph{{Free field representation of
  parafermions and related coset models}},
  \href{https://doi.org/10.1016/0550-3213(89)90224-1}{\emph{Theor. Math. Phys.}
  {\bfseries 83} (1990) 466}.

\bibitem{Kuwahara:1989xy}
M.~Kuwahara, N.~Ohta and H.~Suzuki, \emph{{Conformal field theories realized by
  free fields}},
  \href{https://doi.org/10.1016/0550-3213(90)90454-L}{\emph{Nucl. Phys. B}
  {\bfseries 340} (1990) 448}.

\bibitem{Linshaw:2017tvv}
A.~R. Linshaw, \emph{{Universal two-parameter $\mathcal{W}_{\infty}$-algebra
  and vertex algebras of type $\mathcal{W}(2,3,\dots, N)$}}, {\emph{Compositio
  Mathematica} (2017) } [\href{https://arxiv.org/abs/1710.02275}{{\ttfamily
  1710.02275}}].

\bibitem{Gaberdiel:2012ku}
M.~R. Gaberdiel and R.~Gopakumar, \emph{{Triality in minimal model
  holography}}, \href{https://doi.org/10.1007/JHEP07(2012)127}{\emph{JHEP}
  {\bfseries 07} (2012) 127} [\href{https://arxiv.org/abs/1205.2472}{{\ttfamily
  1205.2472}}].

\bibitem{Prochazka:2014gqa}
T.~s. Procházka, \emph{{Exploring $ {\mathcal{W}}_{\infty } $ in the quadratic
  basis}}, \href{https://doi.org/10.1007/JHEP09(2015)116}{\emph{JHEP}
  {\bfseries 09} (2015) 116} [\href{https://arxiv.org/abs/1411.7697}{{\ttfamily
  1411.7697}}].

\bibitem{Prochazka:2017qum}
T.~Proch\'azka and M.~Rap\v{c}\'ak, \emph{{Webs of W-algebras}},
  \href{https://doi.org/10.1007/JHEP11(2018)109}{\emph{JHEP} {\bfseries 11}
  (2018) 109} [\href{https://arxiv.org/abs/1711.06888}{{\ttfamily
  1711.06888}}].

\bibitem{Prochazka:2015deb}
T.~Proch\'azka, \emph{{$ \mathcal{W} $-symmetry, topological vertex and affine
  Yangian}}, \href{https://doi.org/10.1007/JHEP10(2016)077}{\emph{JHEP}
  {\bfseries 10} (2016) 077}
  [\href{https://arxiv.org/abs/1512.07178}{{\ttfamily 1512.07178}}].

\bibitem{DiFrancesco:1997nk}
P.~Di~Francesco, P.~Mathieu and D.~Senechal, \emph{{Conformal field theory}},
  Graduate Texts in Contemporary Physics. Springer-Verlag, New York, 1997,
  \href{https://doi.org/10.1007/978-1-4612-2256-9}{10.1007/978-1-4612-2256-9}.

\bibitem{Argurio:2000tb}
R.~Argurio, A.~Giveon and A.~Shomer, \emph{{Superstrings on AdS$_3$ and
  symmetric products}},
  \href{https://doi.org/10.1088/1126-6708/2000/12/003}{\emph{JHEP} {\bfseries
  12} (2000) 003} [\href{https://arxiv.org/abs/hep-th/0009242}{{\ttfamily
  hep-th/0009242}}].

\bibitem{Hori:2001ax}
K.~Hori and A.~Kapustin, \emph{{Duality of the fermionic 2-D black hole and
  $\mathcal{N}=2$ Liouville theory as mirror symmetry}},
  \href{https://doi.org/10.1088/1126-6708/2001/08/045}{\emph{JHEP} {\bfseries
  08} (2001) 045} [\href{https://arxiv.org/abs/hep-th/0104202}{{\ttfamily
  hep-th/0104202}}].

\bibitem{Ooguri:1995wj}
H.~Ooguri and C.~Vafa, \emph{{Two-dimensional black hole and singularities of
  CY manifolds}},
  \href{https://doi.org/10.1016/0550-3213(96)00008-9}{\emph{Nucl. Phys. B}
  {\bfseries 463} (1996) 55}
  [\href{https://arxiv.org/abs/hep-th/9511164}{{\ttfamily hep-th/9511164}}].

\bibitem{Giveon:1999zm}
A.~Giveon, D.~Kutasov and O.~Pelc, \emph{{Holography for noncritical
  superstrings}},
  \href{https://doi.org/10.1088/1126-6708/1999/10/035}{\emph{JHEP} {\bfseries
  10} (1999) 035} [\href{https://arxiv.org/abs/hep-th/9907178}{{\ttfamily
  hep-th/9907178}}].

\bibitem{Creutzig:2010bt}
T.~Creutzig, Y.~Hikida and P.~B. R\o{}nne, \emph{{The FZZ duality with
  boundary}}, \href{https://doi.org/10.1007/JHEP09(2011)004}{\emph{JHEP}
  {\bfseries 09} (2011) 004} [\href{https://arxiv.org/abs/1012.4731}{{\ttfamily
  1012.4731}}].

\bibitem{CFL}
T.~Creutzig, B.~Feigin and A.~R. Linshaw, \emph{{$\mathcal{N} = 4$
  superconformal algebras and diagonal cosets}},
  \href{https://doi.org/10.1093/imrn/rnaa078}{\emph{International Mathematics
  Research Notices} (2020) }
  [\href{https://arxiv.org/abs/1910.01228}{{\ttfamily 1910.01228}}].

\bibitem{Gaberdiel:2018rqv}
M.~R. Gaberdiel and R.~Gopakumar, \emph{{Tensionless string spectra on
  AdS$_{3}$}}, \href{https://doi.org/10.1007/JHEP05(2018)085}{\emph{JHEP}
  {\bfseries 05} (2018) 085}
  [\href{https://arxiv.org/abs/1803.04423}{{\ttfamily 1803.04423}}].

\bibitem{Eberhardt:2018ouy}
L.~Eberhardt, M.~R. Gaberdiel and R.~Gopakumar, \emph{{The worldsheet dual of
  the symmetric product CFT}},
  \href{https://doi.org/10.1007/JHEP04(2019)103}{\emph{JHEP} {\bfseries 04}
  (2019) 103} [\href{https://arxiv.org/abs/1812.01007}{{\ttfamily
  1812.01007}}].

\bibitem{Giribet:2018ada}
G.~Giribet, C.~Hull, M.~Kleban, M.~Porrati and E.~Rabinovici,
  \emph{{Superstrings on AdS$_{3}$ at $k= 1$}},
  \href{https://doi.org/10.1007/JHEP08(2018)204}{\emph{JHEP} {\bfseries 08}
  (2018) 204} [\href{https://arxiv.org/abs/1803.04420}{{\ttfamily
  1803.04420}}].

\bibitem{Eberhardt:2019ywk}
L.~Eberhardt, M.~R. Gaberdiel and R.~Gopakumar, \emph{{Deriving the
  AdS$_{3}$/CFT$_{2}$ correspondence}},
  \href{https://doi.org/10.1007/JHEP02(2020)136}{\emph{JHEP} {\bfseries 02}
  (2020) 136} [\href{https://arxiv.org/abs/1911.00378}{{\ttfamily
  1911.00378}}].

\bibitem{Eberhardt:2020akk}
L.~Eberhardt, \emph{{AdS$_{3}$/CFT$_{2}$ at higher genus}},
  \href{https://doi.org/10.1007/JHEP05(2020)150}{\emph{JHEP} {\bfseries 05}
  (2020) 150} [\href{https://arxiv.org/abs/2002.11729}{{\ttfamily
  2002.11729}}].

\bibitem{Hikida:2020kil}
Y.~Hikida and T.~Liu, \emph{{Correlation functions of symmetric orbifold from
  AdS$_3$ string theory}},
  \href{https://doi.org/10.1007/JHEP09(2020)157}{\emph{JHEP} {\bfseries 09}
  (2020) 157} [\href{https://arxiv.org/abs/2005.12511}{{\ttfamily
  2005.12511}}].

\bibitem{Dei:2020zui}
A.~Dei, M.~R. Gaberdiel, R.~Gopakumar and B.~Knighton, \emph{{Free field
  world-sheet correlators for ${\rm AdS}_3$}},
  \href{https://arxiv.org/abs/2009.11306}{{\ttfamily 2009.11306}}.

\bibitem{Gaberdiel:2013vva}
M.~R. Gaberdiel and R.~Gopakumar, \emph{{Large $\mathcal{N}=4$ holography}},
  \href{https://doi.org/10.1007/JHEP09(2013)036}{\emph{JHEP} {\bfseries 09}
  (2013) 036} [\href{https://arxiv.org/abs/1305.4181}{{\ttfamily 1305.4181}}].

\bibitem{Gaberdiel:2014cha}
M.~R. Gaberdiel and R.~Gopakumar, \emph{{Higher spins \& strings}},
  \href{https://doi.org/10.1007/JHEP11(2014)044}{\emph{JHEP} {\bfseries 11}
  (2014) 044} [\href{https://arxiv.org/abs/1406.6103}{{\ttfamily 1406.6103}}].

\bibitem{Creutzig:2013tja}
T.~Creutzig, Y.~Hikida and P.~B. R\o{}nne, \emph{{Extended higher spin
  holography and Grassmannian models}},
  \href{https://doi.org/10.1007/JHEP11(2013)038}{\emph{JHEP} {\bfseries 11}
  (2013) 038} [\href{https://arxiv.org/abs/1306.0466}{{\ttfamily 1306.0466}}].

\bibitem{Creutzig:2014ula}
T.~Creutzig, Y.~Hikida and P.~B. R\o{}nne, \emph{{Higher spin AdS$_{3}$
  holography with extended supersymmetry}},
  \href{https://doi.org/10.1007/JHEP10(2014)163}{\emph{JHEP} {\bfseries 10}
  (2014) 163} [\href{https://arxiv.org/abs/1406.1521}{{\ttfamily 1406.1521}}].

\bibitem{Kazama:1988uz}
Y.~Kazama and H.~Suzuki, \emph{{Characterization of $\mathcal{N}=2$
  superconformal models generated by coset space method}},
  \href{https://doi.org/10.1016/0370-2693(89)91378-6}{\emph{Phys. Lett. B}
  {\bfseries 216} (1989) 112}.

\bibitem{Kazama:1988qp}
Y.~Kazama and H.~Suzuki, \emph{{New $\mathcal{N}=2$ superconformal field
  theories and superstring compactification}},
  \href{https://doi.org/10.1016/0550-3213(89)90250-2}{\emph{Nucl. Phys. B}
  {\bfseries 321} (1989) 232}.

\end{thebibliography}

\providecommand{\href}[2]{#2}\begingroup\raggedright\endgroup

\end{document}